\PassOptionsToPackage{unicode}{hyperref}
\PassOptionsToPackage{hyphens}{url}
\PassOptionsToPackage{dvipsnames,svgnames,x11names}{xcolor}
\documentclass[
  letterpaper,
  DIV=11,
  numbers=noendperiod]{scrartcl}
\usepackage{xcolor}
\usepackage{amsmath,amssymb}
\usepackage{iftex}
\ifPDFTeX
  \usepackage[T1]{fontenc}
  \usepackage[utf8]{inputenc}
  \usepackage{textcomp} 
\else 
  \usepackage{unicode-math} 
  \defaultfontfeatures{Scale=MatchLowercase}
  \defaultfontfeatures[\rmfamily]{Ligatures=TeX,Scale=1}
\fi
\usepackage{lmodern}
\ifPDFTeX\else
\fi
\IfFileExists{upquote.sty}{\usepackage{upquote}}{}
\IfFileExists{microtype.sty}{
  \usepackage[]{microtype}
  \UseMicrotypeSet[protrusion]{basicmath} 
}{}
\makeatletter
\@ifundefined{KOMAClassName}{
  \IfFileExists{parskip.sty}{%
    \usepackage{parskip}
  }{
    \setlength{\parindent}{0pt}
    \setlength{\parskip}{6pt plus 2pt minus 1pt}}
}{
  \KOMAoptions{parskip=half}}
\makeatother
\makeatletter
\ifx\paragraph\undefined\else
  \let\oldparagraph\paragraph
  \renewcommand{\paragraph}{
    \@ifstar
      \xxxParagraphStar
      \xxxParagraphNoStar
  }
  \newcommand{\xxxParagraphStar}[1]{\oldparagraph*{#1}\mbox{}}
  \newcommand{\xxxParagraphNoStar}[1]{\oldparagraph{#1}\mbox{}}
\fi
\ifx\subparagraph\undefined\else
  \let\oldsubparagraph\subparagraph
  \renewcommand{\subparagraph}{
    \@ifstar
      \xxxSubParagraphStar
      \xxxSubParagraphNoStar
  }
  \newcommand{\xxxSubParagraphStar}[1]{\oldsubparagraph*{#1}\mbox{}}
  \newcommand{\xxxSubParagraphNoStar}[1]{\oldsubparagraph{#1}\mbox{}}
\fi
\makeatother

\usepackage{longtable,booktabs,array}
\usepackage{calc} 
\usepackage{etoolbox}
\makeatletter
\patchcmd\longtable{\par}{\if@noskipsec\mbox{}\fi\par}{}{}
\makeatother
\IfFileExists{footnotehyper.sty}{\usepackage{footnotehyper}}{\usepackage{footnote}}
\makesavenoteenv{longtable}
\usepackage{graphicx}
\usepackage{pdflscape}
\makeatletter
\newsavebox\pandoc@box
\newcommand*\pandocbounded[1]{
  \sbox\pandoc@box{#1}%
  \Gscale@div\@tempa{\textheight}{\dimexpr\ht\pandoc@box+\dp\pandoc@box\relax}%
  \Gscale@div\@tempb{\linewidth}{\wd\pandoc@box}%
  \ifdim\@tempb\p@<\@tempa\p@\let\@tempa\@tempb\fi
  \ifdim\@tempa\p@<\p@\scalebox{\@tempa}{\usebox\pandoc@box}%
  \else\usebox{\pandoc@box}%
  \fi%
}
\def\fps@figure{htbp}
\makeatother

\NewDocumentCommand\citeproctext{}{}

\makeatletter
 \let\@cite@ofmt\@firstofone
 \def\@biblabel#1{}
 \def\@cite#1#2{{#1\if@tempswa , #2\fi}}
\makeatother
\newlength{\cslhangindent}
\newlength{\csllabelwidth}
\newenvironment{CSLReferences}[2] 
 {\begin{list}{}{%
  \setlength{\itemindent}{0pt}
  \setlength{\leftmargin}{0pt}
  \setlength{\parsep}{0pt}
  \ifodd #1
   \setlength{\leftmargin}{\cslhangindent}
   \setlength{\itemindent}{-1\cslhangindent}
  \fi
  \setlength{\itemsep}{#2\baselineskip}}}
 {\end{list}}
\usepackage{calc}

\providecommand{\tightlist}{%
  \setlength{\itemsep}{0pt}\setlength{\parskip}{0pt}}

\usepackage{booktabs}
\usepackage{caption}
\usepackage{longtable}
\usepackage{colortbl}
\usepackage{array}
\usepackage{anyfontsize}
\usepackage{multirow}
\KOMAoption{captions}{tableheading}
\makeatletter
\@ifpackageloaded{caption}{}{\usepackage{caption}}
\AtBeginDocument{%
\ifdefined\contentsname
  \renewcommand*\contentsname{Table of contents}
\else
  \newcommand\contentsname{Table of contents}
\fi
\ifdefined\listfigurename
  \renewcommand*\listfigurename{List of Figures}
\else
  \newcommand\listfigurename{List of Figures}
\fi
\ifdefined\listtablename
  \renewcommand*\listtablename{List of Tables}
\else
  \newcommand\listtablename{List of Tables}
\fi
\ifdefined\figurename
  \renewcommand*\figurename{Figure}
\else
  \newcommand\figurename{Figure}
\fi
\ifdefined\tablename
  \renewcommand*\tablename{Table}
\else
  \newcommand\tablename{Table}
\fi
}
\@ifpackageloaded{float}{}{\usepackage{float}}
\floatstyle{ruled}
\@ifundefined{c@chapter}{\newfloat{codelisting}{h}{lop}}{\newfloat{codelisting}{h}{lop}[chapter]}
\floatname{codelisting}{Listing}

\makeatother
\makeatletter
\@ifpackageloaded{caption}{}{\usepackage{caption}}
\@ifpackageloaded{subcaption}{}{\usepackage{subcaption}}
\makeatother
\usepackage{bookmark}
\IfFileExists{xurl.sty}{\usepackage{xurl}}{} 
\hypersetup{
  pdftitle={PIONEER: Bayesian Joint Modelling of Mechanistic Tumour Growth and Time-to-Event Endpoints for Dynamic Prediction of Ongoing Oncology Trials},
  pdfauthor={Karim Naguib; Roger Berché; Lu Li; Antonia Bevan; Sajan Khosla; Jessica Davies; Paul Metcalfe},
  pdfkeywords={Bayesian joint modelling, tumour dynamics, multistate
survival, dynamic prediction, oncology trials, progression-free
survival},
  colorlinks=true,
  linkcolor={blue},
  filecolor={Maroon},
  citecolor={Blue},
  urlcolor={Blue},
  pdfcreator={LaTeX via pandoc}}

\title{PIONEER: Bayesian Joint Modelling of Mechanistic Tumour Growth
and Time-to-Event Endpoints for Dynamic Prediction of Ongoing Oncology
Trials}
\author{Karim Naguib \and Roger Berché \and Lu Li \and Antonia
Bevan \and Sajan Khosla \and Jessica Davies \and Paul Metcalfe}
\date{2026-06-26}
\begin{document}
\maketitle
\begin{abstract}
High-stakes decisions in oncology clinical trials must often be made
while survival data remains immature: progression-free survival (PFS)
and overall survival (OS) are heavily censored, few events have
accumulated, and the primary endpoint may be months or years from
reading out. What is available at interim data cut-offs is
information-rich longitudinal tumour measurements and baseline
covariates. We present PIONEER, a Bayesian joint modelling framework
that couples a mechanistic two-component state-space submodel of
longitudinal tumour size dynamics to a multistate proportional-hazard
submodel for competing clinical events, fitted simultaneously under a
single posterior. The mechanistic submodel infers latent per-patient
tumour trajectories --- decomposed into treatment-responsive and
refractory compartments with Gompertz-attenuated growth --- from sparse,
noisy sum-of-longest-diameter (SLD) observations. These latent
trajectories feed the multistate hazard as time-varying covariates,
while the event data simultaneously refines the tumour dynamics through
the joint likelihood. All clinical endpoints (PFS, OS, objective
response rate) are derived from the joint posterior in a single
forward-simulation pass, propagating full parameter uncertainty without
any two-stage plug-in. Applied to a case study in extensive-stage
small-cell lung cancer (two trials, \(N = 497\)), leave-future-out
cross-validation demonstrates that at month 4 of enrolment (9 patients)
the model produces calibrated PFS forecasts covering the mature month-19
Kaplan--Meier curve, and at month 11 (39 patients) the OS forecast
converges --- representing at least 8 months of advance forecasting with
properly quantified uncertainty. We hope this work paves the way for
broader adoption of Bayesian mechanistic state-space frameworks in
clinical development, enabling earlier and more informed decision-making
from immature trial data.
\end{abstract}

\section{Introduction}\label{sec-introduction}

Oncology drug development relies on early-phase (Phase 1/2) trials to
establish safety and preliminary evidence of anti-tumour activity, with
long-term efficacy endpoints, such as recurrence-free survival (RFS),
progression-free survival (PFS), and overall survival (OS), ultimately
serving as the primary basis for Phase 3 decision-making and regulatory
approval. Yet despite their central role, these endpoints are frequently
immature at the point when critical go/no-go decisions must be made,
forcing development teams to act under substantial uncertainty. The
consequences of this are significant: costly late-stage failures,
prolonged and resource-intensive trials, and meaningful delays to
patients accessing potentially transformative regimens. What is needed
is a principled approach to extract the maximum signal from the data
that is available, one that can translate early longitudinal
measurements into reliable projections of long-term clinical outcomes,
contextualize those projections against standard-of-care benchmarks, and
quantify the uncertainty inherent in doing so.

A substantial body of work has addressed this problem from different
angles. The starting point for any longitudinal-tumour-to-survival
framework is the empirical observation that tumour size kinetics
correlate strongly with patient survival, and that this correlation is
captured well by a two-exponential ``decrease + growth'' decomposition.
Stein and colleagues introduced the formulation in advanced prostate
cancer (Stein et al. 2008, 2011), showing that the growth-rate constant
in particular is a strong correlate of OS. Wang et al. (2009) extended
the framework to non-small-cell lung cancer (NSCLC). Claret and
co-authors (Claret et al. 2009, 2012) turned the framework into a
forecasting tool, demonstrating that Phase II tumour dynamics can be
combined with a parametric OS submodel to predict Phase III OS, with
calibration successes in colorectal and NSCLC contexts. Bruno et al.
(2014) consolidated this body of work into a methodological review, and
Bruno et al. (2020) argued explicitly for tumour-dynamic modelling as
the standard supporting tool for go/no-go and dose-decision use-cases.
Tumour-growth-inhibition (TGI) modelling --- typically the
decrease/growth two-exponential framework above, fitted to Phase II SLD
data and used to predict Phase III OS --- is now the dominant
longitudinal-tumour-to-survival paradigm in pharmacometrics. These early
implementations follow a two-stage workflow: fit dynamics first, then
use derived summary statistics (e.g.~growth rate, time to nadir, week-8
SLD) as covariates in a downstream survival model.

The natural starting point for survival analysis is the Cox
proportional-hazards model, fitted with treatment arm and baseline
covariates. Extending it to incorporate a longitudinal biomarker like
SLD requires the time-varying-covariate form. Doing so correctly is,
however, non-trivial for two reasons. First, the observed SLD is a noisy
measurement of the patient's true tumour trajectory; plugging in the
observed values as a covariate attenuates the estimated SLD--hazard
association in proportion to the measurement noise --- a direct
errors-in-variables problem. Second, a patient's unmeasured
characteristics --- intrinsic disease aggressiveness, treatment
sensitivity, immune competence --- jointly determine both their tumour
dynamics and their risk of progression and death. Any model that omits
this shared latent structure is misspecified: the SLD covariate absorbs
variation that rightly belongs to the unmeasured individual, and the
resulting coefficient is inconsistent. The two-stage TGI workflow
compounds both problems --- summary statistics derived from a
first-stage kinetic fit are treated as fixed inputs in the survival
stage, inheriting the measurement noise of the SLD data and discarding
uncertainty about the kinetic parameters themselves. The principled fix
is a \emph{joint longitudinal-survival} model: introduce the biomarker's
underlying latent trajectory as a shared patient-level latent variable,
and let both the longitudinal observations and the survival hazard
depend on it simultaneously. The hazard then conditions on the smooth,
noise-free latent state; the shared patient parameters are informed by
both the SLD record and the event history at once; and the
misspecification from omitting unmeasured individual heterogeneity is
eliminated by construction. The textbook treatment is Rizopoulos (2012),
with Hickey et al. (2016) reviewing extensions to multiple longitudinal
outcomes and competing risks.

A handful of TGI papers have adopted this joint-model architecture
explicitly. These models retain the hierarchical (mixed-effects) tumour
submodel --- each patient has their own kinetic parameters drawn from a
population distribution --- but couple it to the survival submodel
through the shared latent trajectory rather than through plug-in
summaries. Ribba et al. (2014) reviews the underlying mixed-effects
methodology across the Simeoni (Simeoni et al. 2004), Claret (Claret et
al. 2009), Stein (Stein et al. 2008) and Wang (Wang et al. 2009)
families. Tardivon et al. (2019) fit a population NLME tumour-size model
jointly with an OS submodel in atezolizumab-treated urothelial
carcinoma. Desmée et al. (2017) introduce Hamiltonian Monte Carlo
Bayesian inference for the same joint structure in metastatic prostate
cancer. In the immuno-oncology setting specifically, Kerioui et al.
(2020) develop a Bayesian HMC joint model with a tumour-size submodel
feeding an OS hazard; Kerioui et al. (2022) provides a non-technical
introduction.

Multistate (progression-death) extensions of competing-risks survival
have been used extensively in chronic-disease epidemiology and,
increasingly, in oncology, where post-progression survival is itself of
clinical interest (Putter et al. 2007; Meira-Machado et al. 2009). The
framework is sometimes called \emph{illness-death} in biostatistics,
although in oncology trials all patients are already ill at enrolment
and the relevant transition is from controlled to progressive disease.
Modelling progression and death as separate transitions has two
practical advantages over a single OS hazard: it lets the data speak to
clinically distinct mechanisms --- early progression, direct death
without prior progression, and post-progression death have different
time-courses and different covariate dependencies --- and it enables
forecasting the full PFS--OS joint distribution rather than each
endpoint in isolation.

Despite the maturity of each strand above, several gaps remain. First,
existing tumour-dynamics-to-survival frameworks feed the dynamics into a
\emph{single} hazard (typically OS); they cannot express that tumour
features affect progression, death without progression, and
post-progression death differently --- yet these are clinically distinct
mechanisms. Second, most published work fits the tumour model first,
then plugs the fitted trajectories into a survival model; this two-stage
approach ignores the feedback from event data to the dynamics and
underestimates uncertainty. Third, trial-level pooling --- sharing
information across historical and ongoing trials rather than handling
each independently via ad-hoc meta-analysis --- is rarely built into the
framework as a first-class component. Fourth, endpoints (PFS, OS, ORR)
are typically computed from point estimates of upstream parameters
rather than propagated through the full posterior.

We present PIONEER\footnote{Predictive Intelligence for ONcology Early
  Evaluation and Response.}, a framework built to close these gaps. The
model couples a \emph{mechanistic} two-component state-space submodel of
longitudinal tumour size (sum of longest diameters, SLD) to a
\emph{multistate proportional-hazard} submodel for the clinical events
of interest inside a single Bayesian hierarchical framework. All
clinical endpoints --- target lesion progression, multistate PFS, OS,
objective response rate --- are derived in the same posterior-predictive
pass from the joint posterior, so every reported quantity automatically
inherits the model's full uncertainty. The mechanistic submodel feeds
the multistate submodel through a shared \emph{latent} trajectory: the
latent SLD, decrease rate and growth rate enter the multistate hazards
as time-varying covariates, and because both submodels depend on the
same patient-level latent parameters, the event data refines the
inferred trajectory at the same time as the SLD data does. There is no
two-stage plug-in.

The novelty has five facets:

\begin{itemize}
\tightlist
\item
  \textbf{Transition-specific tumour covariates.} Each transition in the
  illness-death model has its own tumour-derived covariate vector with
  transition-specific coefficients, allowing the model to express that
  growth rate strongly predicts progression hazard but only weakly
  affects post-progression mortality --- a distinction single-hazard
  joint models cannot make.
\item
  \textbf{Joint latent state.} The decrease/growth decomposition is the
  latent state of a state-space model fitted jointly with the multistate
  submodel --- not a two-stage plug-in --- avoiding both the attenuation
  and the omitted-frailty misspecification of plug-in approaches and
  correctly propagating kinetic uncertainty into every downstream
  endpoint.
\item
  \textbf{Correlated per-transition frailty.} Each transition carries
  its own patient-level frailty, and the frailty terms are modelled
  jointly as a multivariate normal with an estimated correlation
  structure, capturing residual individual heterogeneity unexplained by
  the tumour-derived time-varying covariates and revealing
  cross-transition dependence in unmeasured patient characteristics.
\item
  \textbf{Multi-trial hierarchy.} Patients are nested within arms within
  trials in a single hierarchical model; population-level parameters are
  shared across trials by construction, so fitting on multiple trials
  simultaneously yields cross-trial borrowing without a separate
  meta-analytic step. Covariate effects are entered via QR-decomposed
  design matrices with elicited priors mapped through the rotation.
\item
  \textbf{Full-posterior endpoint propagation.} All endpoints are read
  off the joint posterior, producing both unconditional
  posterior-predictive and observed-data-conditioned predictive
  distributions without re-fitting.
\end{itemize}

PIONEER is built around a \emph{backward inference, forward simulation}
pipeline. The backward step performs joint Bayesian inference over the
entire trial --- latent tumour trajectories, multistate hazards and the
population hierarchy, all in a single posterior --- using every
patient's observed SLD record and event history through the data
cut-off. The forward step then propagates each posterior draw past the
cut-off, extending the latent trajectory, accumulating the multistate
hazards, and producing posterior predictive distributions for mature-DCO
Kaplan--Meier curves, response rates, and any other endpoint of
interest. Inference and forecasting share the same parameter uncertainty
by construction; there is no re-fitting and no plug-in step between
them. The credibility of the forward simulation rests on two pillars:
the mechanistic parametric structure of the tumour submodel, whose
decrease/growth decomposition constrains the shape of the extrapolated
trajectory; and the population hierarchy, which --- when historical
trials are included --- encodes what kinetic profiles and hazard
trajectories tend to look like at maturities the target trial has not
yet reached.

\section{Model}\label{sec-model}

\begin{figure}

\centering{

\includegraphics[width=1\linewidth,height=\textheight,keepaspectratio]{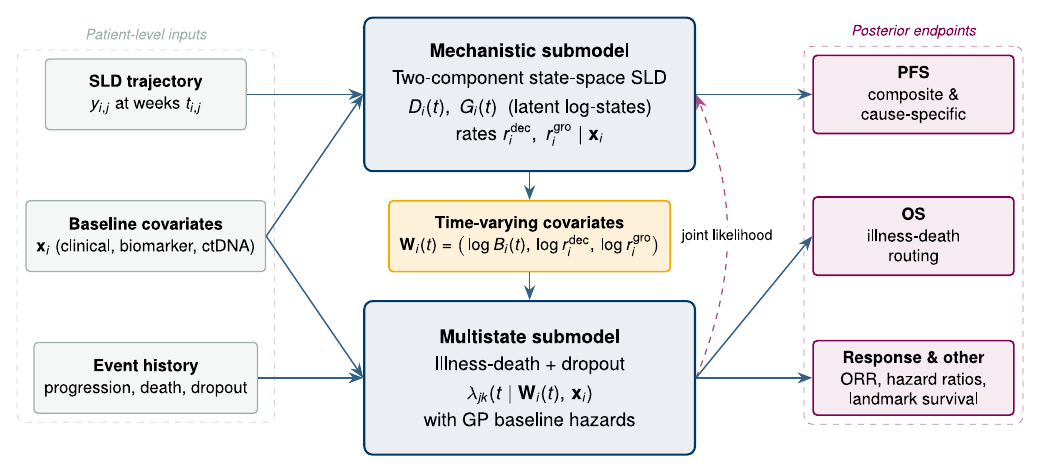}

}

\caption{\label{fig-model-overview}\textbf{Joint model overview.}
Patient-level inputs (left) feed two coupled submodels (centre): the
mechanistic state-space submodel of SLD dynamics, and the multistate
illness-death-with-dropout hazard submodel. The tumour-derived
time-varying covariate vector \(\mathbf{W}_i(t)\) is the explicit
\emph{bridge} through which the mechanistic submodel feeds the hazards;
the dashed arrow indicates the joint likelihood, which lets the event
data refine the latent state's posterior. Endpoints (right) are
deterministic functions of the joint posterior.}

\end{figure}

This section specifies the joint model in full. We organise the
description around the data flow shown in
Figure~\ref{fig-model-overview}: patient-level inputs (longitudinal SLD,
baseline covariates, event histories) are routed through two coupled
submodels --- a mechanistic state-space submodel for the SLD trajectory
(Section~\ref{sec-mechanistic}) and a multistate hazard submodel for the
clinical events (Section~\ref{sec-multistate}) --- linked by a small set
of time-varying tumour-derived covariates (the \emph{bridge}). All
clinical endpoints are read off the joint posterior in a single
posterior-predictive pass, with no two-stage plug-in.

\subsection{Notation and setup}\label{sec-notation}

We index patients by \(i = 1, \dots, N\). The time clock is the
patient's \emph{study week}, an integer-valued grid in which week \(1\)
denotes the first week on treatment; non-positive weeks (\(t \le 0\))
correspond to the pre-treatment screening window. Each patient \(i\)
contributes a vector of baseline covariates
\(\mathbf{x}_i \in \mathbb{R}^p\), a set of SLD assessments at visit
weeks \(\mathcal{V}_i \subset \mathbb{Z}_{>0}\) (ragged across
patients), and an event history. We will write \(y_i(t)\) for the SLD
measurement at visit week \(t \in \mathcal{V}_i\) --- a single scalar
per visit --- so that \(y_i(\cdot)\) is a function defined only on the
patient's visit weeks. The event history takes one of four mutually
exclusive terminal states by the data cut-off: alive and
progression-free (state \(0\)), progressed (state \(1\)), dead (state
\(2\)), or off-trial without death (state \(3\)). Transitions between
these states are written \(j \to k\); the five admissible transitions
are \(0 \to 1\), \(0 \to 2\), \(1 \to 2\), \(0 \to 3\) and \(3 \to 2\)
(see Figure~\ref{fig-multistate-diagram}).

For each patient we identify a single \emph{baseline visit}, defined as
the last SLD assessment in the pre-treatment screening window. We write
\(t_i^{\mathrm{bl}} \le 0\) for the week of patient \(i\)'s baseline
visit, and \(y_i^{\mathrm{bl}} \equiv y_i(t_i^{\mathrm{bl}})\) for the
SLD measurement at that visit. We define the \emph{normalised burden}
using the patient's \emph{latent} SLD trajectory \(\mathrm{SLD}_i(t)\)
--- the true, unobserved tumour size at week \(t\) that underlies the
noisy visit measurements \(y_i(t)\), \(t \in \mathcal{V}_i\), and is
modelled explicitly in Section~\ref{sec-mechanistic}:
\begin{equation}\protect\phantomsection\label{eq-sld-normalised}{
B_i(t) \;=\; \frac{\mathrm{SLD}_i(t)}{y_i^{\mathrm{bl}}},
}\end{equation} so that the latent burden at the baseline visit,
\(B_i(t_i^{\mathrm{bl}})\), is the ratio of the latent SLD to the
observed baseline SLD at that visit; up to baseline measurement noise,
\(B_i(t_i^{\mathrm{bl}}) \approx 1\). The symbol \(B\) is a mnemonic for
\emph{burden}; values of \(B_i(t)\) above \(1\) correspond to a tumour
size larger than baseline, values below \(1\) to a tumour size smaller
than baseline. The observed baseline \(y_i^{\mathrm{bl}}\) enters only
as a known offset on the normaliser --- we \emph{condition} on it as the
patient's reference scale rather than modelling it as a noisy outcome of
the dynamics --- so the model is invariant to overall tumour-size scale
and the dynamics are not confounded with cohort-level differences in
initial disease burden. The same definition generalises naturally to
settings where SLD is replaced by another biomarker (e.g., PSA in
prostate cancer).

Trial membership is denoted \(s[i] \in \{1, \dots, S\}\) when relevant,
where \(S\) is the number of trials in the analysis.

\subsection{Mechanistic tumour-dynamics submodel}\label{sec-mechanistic}

\subsubsection{Latent two-component
representation}\label{sec-mech-states}

We model the normalised burden \(B_i(t)\) as the sum of two latent
components --- a \emph{decreasing} component \(D_i(t) > 0\) representing
burden that responds to treatment, and a \emph{growing} component
\(G_i(t) > 0\) representing burden that is refractory or
post-progression:
\begin{equation}\protect\phantomsection\label{eq-sld-decomp}{
B_i(t) \;=\; D_i(t) \;+\; G_i(t).
}\end{equation} We emphasise that this is a \emph{phenomenological}
decomposition: \(D_i\) and \(G_i\) are not claimed to recover real
clonal subpopulations. The interpretation is operational --- \(D_i\)
tracks the part of the burden that contracts under therapy and \(G_i\)
the part that expands. The case for the decomposition is forecasting
performance and parameter transportability, not a mechanistic biological
claim.

Like \(B_i(t)\) itself, \(D_i(t)\) and \(G_i(t)\) are dimensionless
contributions to the normalised burden, and at baseline they sum to one.
The dynamics of \(D_i\) and \(G_i\) --- exponential decay and
exponential growth, respectively --- are most naturally written on the
log scale, and inference is performed in terms of \(\log D_i\) and
\(\log G_i\) (next subsection); we keep the \emph{names} \(D_i, G_i\) on
the linear scale to match \(B_i\) and write logs explicitly where they
apply.

\subsubsection{Dynamics}\label{sec-mech-dynamics}

The two components evolve via patient-specific \emph{change rates}: a
positive rate of contraction \(r^{\mathrm{dec}}_i\) for the decreasing
component and a positive rate of expansion \(r^{\mathrm{gro}}_i\) for
the growing component:
\begin{equation}\protect\phantomsection\label{eq-mech-ode}{
\begin{aligned}
\frac{\mathrm{d}\log D_i(t)}{\mathrm{d} t} &\;=\; -\, r^{\mathrm{dec}}_i, \\
\frac{\mathrm{d}\log G_i(t)}{\mathrm{d} t} &\;=\; +\, r^{\mathrm{gro}}_i,
\end{aligned}
}\end{equation} i.e.~exponential decay of \(D_i(t)\) at rate
\(r^{\mathrm{dec}}_i\) and exponential growth of \(G_i(t)\) at rate
\(r^{\mathrm{gro}}_i\). Substituting (Equation~\ref{eq-mech-discrete})
into (Equation~\ref{eq-sld-decomp}) gives the closed-form trajectory
\begin{equation}\protect\phantomsection\label{eq-sld-twoexp}{
B_i(t)
\;=\;
\pi_i\,\mathrm{e}^{-r^{\mathrm{dec}}_i\,(t - t_i^{\mathrm{bl}})}
\;+\;
(1 - \pi_i)\,\mathrm{e}^{+r^{\mathrm{gro}}_i\,(t - t_i^{\mathrm{bl}})},
}\end{equation} where \(\pi_i\) is the patient's initial decreasing
fraction. This is the familiar two-exponential SLD trajectory of the
Stein/Claret/Bruno tradition (Stein et al. 2008); inference is performed
on \(\log D_i, \log G_i\) rather than on (Equation~\ref{eq-sld-twoexp})
as a regression target.

In the primary configuration, the \emph{decrease} rate
\(r^{\mathrm{dec}}_i\) is constant in time and patient-specific; the
\emph{growth} rate is modulated by a Gompertz decay term (see
Section~\ref{sec-mech-gompertz}). Both rates depend on baseline
covariates through their decomposition into a magnitude and a balance,
described next.\footnote{The framework supports a natural extension in
  which the constant rates of (Equation~\ref{eq-mech-discrete}) are
  replaced with smoothly time-varying versions, generated by adding a
  stationary AR(1) process to each log-rate at the population (or,
  optionally, patient) level. The AR(1) mean is the patient's
  \(\log r^{\mathrm{dec}}_i\) or \(\log r^{\mathrm{gro}}_i\) from
  (Equation~\ref{eq-rate-pair}); the AR(1) innovations capture smooth
  deviations of the realised rate from its time-averaged value over the
  course of follow-up. We do not exercise this option in the primary
  analysis because (i) the observation density already accommodates
  measurement error and (ii) the smoothed posterior trajectories from
  the constant-rate model are visibly close to the observed SLD curves
  in the case study.}

The discretisation we use for inference puts \(\log D_i\) and
\(\log G_i\) on a weekly grid running from \(t_i^{\mathrm{bl}}\) to a
patient-specific horizon \(T_i\):
\begin{equation}\protect\phantomsection\label{eq-mech-discrete}{
\begin{aligned}
\log D_i(t) &\;=\; \log D_i(t_i^{\mathrm{bl}}) - r^{\mathrm{dec}}_i \cdot (t - t_i^{\mathrm{bl}}), \\
\log G_i(t) &\;=\; \log G_i(t_i^{\mathrm{bl}}) + r^{\mathrm{gro}}_i \cdot (t - t_i^{\mathrm{bl}}),
\end{aligned}
}\end{equation}

i.e.~the log-components are linear in elapsed time since baseline when
the rates are constant; equivalently, \(D_i\) and \(G_i\) themselves are
exponential in \(t - t_i^{\mathrm{bl}}\). The full latent trajectory
\(\{(D_i(t), G_i(t))\}_{t \ge t_i^{\mathrm{bl}}}\) is therefore
determined by three patient-level quantities --- the initial split
between the two components at the baseline visit and the two change
rates --- plus the observed baseline measurement \(y_i^{\mathrm{bl}}\).

\subsubsection{Gompertz growth-rate decay}\label{sec-mech-gompertz}

A known limitation of the constant-growth-rate form in
(Equation~\ref{eq-mech-discrete}) is that \(G_i(t)\) grows without bound
as \(t \to \infty\): the forecast burden accumulates linearly in
log-space at rate \(r^{\mathrm{gro}}_i\) forever, producing biologically
implausible trajectories at long horizons (Bruno et al. 2014, 2023). The
practical consequence in leave-future-out validation on the SCLC cohort
is a systematic over-prediction of re-growth at late forecast windows: a
deterministic bi-exponential crosses the RECIST progression threshold at
predictable long-horizon times regardless of whether the data actually
show such acceleration (see Section~\ref{sec-results}).

The pharmacometric literature addresses this through \textbf{Gompertz
growth-rate decay}: rather than treating \(r^{\mathrm{gro}}_i\) as a
constant, the instantaneous specific growth rate of the growing
compartment is allowed to decay exponentially in time,
\begin{equation}\protect\phantomsection\label{eq-gompertz-rate}{
g_i(t) \;=\; r^{\mathrm{gro}}_i \cdot \mathrm{e}^{-\kappa_i\,(t - t_i^{\mathrm{bl}})},
}\end{equation} where \(\kappa_i > 0\) is a patient-level Gompertz decay
parameter. The decrease compartment is unaffected --- it already
contracts to zero and does not contribute to long-horizon explosion.
Integrating (Equation~\ref{eq-gompertz-rate}) from baseline gives the
\emph{warped growth time} as a function of elapsed time
\(\Delta t = t - t_i^{\mathrm{bl}}\):
\begin{equation}\protect\phantomsection\label{eq-gompertz-warp}{
\varphi_i(\Delta t) \;=\; \int_0^{\Delta t} \mathrm{e}^{-\kappa_i\,u}\,\mathrm{d}u
\;=\; \frac{1 - \mathrm{e}^{-\kappa_i\,\Delta t}}{\kappa_i}.
}\end{equation} Note that in the base model
(Equation~\ref{eq-mech-discrete}), \(\varphi_i(\Delta t) = \Delta t\)
--- Gompertz simply replaces linear elapsed time with this saturating
warp, and the base case is recovered exactly as \(\kappa_i \to 0\). The
warped time \(\varphi_i(t - t_i^{\mathrm{bl}})\) replaces the elapsed
time \(t - t_i^{\mathrm{bl}}\) in the growth arm of
(Equation~\ref{eq-mech-discrete}):
\begin{equation}\protect\phantomsection\label{eq-gompertz-discrete}{
\log G_i(t) \;=\; \log G_i(t_i^{\mathrm{bl}}) \;+\; r^{\mathrm{gro}}_i \cdot \varphi_i(t - t_i^{\mathrm{bl}}).
}\end{equation}

The substitution has two important limiting properties: \[
\lim_{\kappa_i \to 0} \varphi_i(\Delta t) \;=\; \Delta t, \qquad
\lim_{\Delta t \to \infty} \varphi_i(\Delta t) \;=\; \frac{1}{\kappa_i},
\] so that \(\kappa_i \to 0\) recovers the constant-rate model
(Equation~\ref{eq-mech-discrete}) exactly, while finite \(\kappa_i > 0\)
causes the log-growth arm to plateau at
\(\log G_i(t_i^{\mathrm{bl}}) + r^{\mathrm{gro}}_i / \kappa_i\) as the
horizon grows --- a finite carrying capacity. The closed-form trajectory
(Equation~\ref{eq-sld-twoexp}) becomes
\begin{equation}\protect\phantomsection\label{eq-sld-gompertz}{
B_i(t)
\;=\;
\pi_i\,\mathrm{e}^{-r^{\mathrm{dec}}_i\,(t - t_i^{\mathrm{bl}})}
\;+\;
(1 - \pi_i)\,\mathrm{e}^{+r^{\mathrm{gro}}_i\,\varphi_i(t - t_i^{\mathrm{bl}})}.
}\end{equation}

The decay parameter is modelled on the log scale with a population-level
intercept:
\begin{equation}\protect\phantomsection\label{eq-kappa-linpred}{
\log \kappa_i \;=\; \alpha^{\kappa},
}\end{equation} where \(\alpha^{\kappa}\) is the population log-decay
intercept, so that \(\kappa_i = \mathrm{e}^{\alpha^{\kappa}} > 0\) for
all patients. Baseline-covariate slopes on \(\log \kappa_i\) are
supported by the model but not activated in the primary analysis; the
current configuration pools all patients at a single population decay
rate. The prior on \(\alpha^{\kappa}\) is weakly informative, chosen so
that the implied carrying capacity \(r^{\mathrm{gro}}_i / \kappa_i\) is
finite and reachable within the forecast window, while remaining diffuse
enough that the posterior can contract toward the curvature actually
present in the observed trajectories.

\subsubsection{Initial state}\label{sec-mech-initial}

By construction, \(B_i(t_i^{\mathrm{bl}}) = 1\). The initial allocation
between the two components is parameterised by the \emph{initial
decreasing fraction} \(\pi_i \in (0, 1)\):
\begin{equation}\protect\phantomsection\label{eq-mech-initial-split}{
D_i(t_i^{\mathrm{bl}}) = \pi_i, \qquad G_i(t_i^{\mathrm{bl}}) = 1 - \pi_i.
}\end{equation} which satisfies
\(D_i(t_i^{\mathrm{bl}}) + G_i(t_i^{\mathrm{bl}}) = 1\) as required by
(Equation~\ref{eq-sld-decomp}). On the log scale used for inference,
this reads \(\log D_i(t_i^{\mathrm{bl}}) = \log \pi_i\) and
\(\log G_i(t_i^{\mathrm{bl}}) = \log(1-\pi_i)\).

We work with \(\pi_i\) on the logit scale,
\begin{equation}\protect\phantomsection\label{eq-init-linpred}{
\eta^{\mathrm{init}}_i \;=\; \mathrm{logit}(\pi_i)
\;=\; \alpha^{\mathrm{init}} + \alpha^{\mathrm{init}}_{s[i]} + \alpha^{\mathrm{init}}_i + \mathbf{x}_i \cdot \boldsymbol{\beta}^{\mathrm{init}},
}\end{equation} where \(\alpha^{\mathrm{init}}\) is a population-level
intercept,
\(\alpha^{\mathrm{init}}_{s[i]} \sim \mathtt{Normal}(0, \tau^{\mathrm{init}}_{\mathrm{arm}}{}^2)\)
a trial-arm deviation,
\(\alpha^{\mathrm{init}}_i \sim \mathtt{Normal}(0, \tau^{\mathrm{init}}_{\mathrm{patient}}{}^2)\)
a patient-level deviation, and \(\boldsymbol{\beta}^{\mathrm{init}}\) a
vector of population-level covariate coefficients. Large positive
\(\eta^{\mathrm{init}}_i\) corresponds to a patient who starts almost
entirely in the responsive compartment (and therefore exhibits a large
early decrease before any growing-component contribution becomes
visible); large negative \(\eta^{\mathrm{init}}_i\) corresponds to a
patient whose SLD trajectory is dominated by growth from week \(1\).

\subsubsection{Patient-level change rates}\label{sec-mech-rates}

Rather than parameterise \(r^{\mathrm{dec}}_i\) and
\(r^{\mathrm{gro}}_i\) directly, we decompose each patient's rate pair
into a \emph{magnitude} (total log-rate \(\eta^{\mathrm{tot}}_i\)) and a
\emph{balance} (\(\eta^{\mathrm{bal}}_i\), logit-scale allocation
between decrease and growth):
\begin{equation}\protect\phantomsection\label{eq-rate-pair}{
\begin{aligned}
\log r^{\mathrm{dec}}_i &\;=\; \eta^{\mathrm{tot}}_i \;+\; \log \sigma(\eta^{\mathrm{bal}}_i), \\
\log r^{\mathrm{gro}}_i &\;=\; \eta^{\mathrm{tot}}_i \;+\; \log\!\big(1 - \sigma(\eta^{\mathrm{bal}}_i)\big),
\end{aligned}
}\end{equation} where \(\sigma(\cdot)\) is the standard logistic
function. Equivalently, the \emph{decreasing fraction of the rate
magnitude} is \(\sigma(\eta^{\mathrm{bal}}_i) \in (0,1)\), and
\(1 - \sigma(\eta^{\mathrm{bal}}_i)\) is the growing fraction; large
positive \(\eta^{\mathrm{bal}}_i\) describes a patient whose dynamics
are dominated by contraction, large negative \(\eta^{\mathrm{bal}}_i\) a
patient whose dynamics are dominated by expansion. The decomposition is
bijective: given \(r^{\mathrm{dec}}_i, r^{\mathrm{gro}}_i\) we recover
\[
\begin{aligned}
\eta^{\mathrm{tot}}_i &\;=\; \log(r^{\mathrm{dec}}_i + r^{\mathrm{gro}}_i), \\
\eta^{\mathrm{bal}}_i &\;=\; \log(r^{\mathrm{dec}}_i / r^{\mathrm{gro}}_i).
\end{aligned}
\]

The magnitude and balance parameters carry the population-, covariate-,
and patient-level structure. We let
\begin{equation}\protect\phantomsection\label{eq-rate-linpred}{
\begin{aligned}
\eta^{\mathrm{tot}}_i &\;=\; \alpha^{\mathrm{tot}} \;+\; \alpha^{\mathrm{tot}}_{s[i]} \;+\; \alpha^{\mathrm{tot}}_i, \\
\eta^{\mathrm{bal}}_i &\;=\; \alpha^{\mathrm{bal}} \;+\; \alpha^{\mathrm{bal}}_{s[i]} \;+\; \alpha^{\mathrm{bal}}_i \;+\; \mathbf{x}_i \cdot \boldsymbol{\beta}^{\mathrm{bal}},
\end{aligned}
}\end{equation} with population-level intercepts
\(\alpha^{\mathrm{tot}}, \alpha^{\mathrm{bal}}\), trial-arm deviations
\(\alpha^{\mathrm{tot}}_{s[i]} \sim \mathtt{Normal}(0, \tau^{\mathrm{tot}}_{\mathrm{arm}}{}^2)\)
and
\(\alpha^{\mathrm{bal}}_{s[i]} \sim \mathtt{Normal}(0, \tau^{\mathrm{bal}}_{\mathrm{arm}}{}^2)\),
patient-level deviations
\(\alpha^{\mathrm{tot}}_i \sim \mathtt{Normal}(0, \tau^{\mathrm{tot}}_{\mathrm{patient}}{}^2)\)
and
\(\alpha^{\mathrm{bal}}_i \sim \mathtt{Normal}(0, \tau^{\mathrm{bal}}_{\mathrm{patient}}{}^2)\),
and a covariate coefficient vector \(\boldsymbol{\beta}^{\mathrm{bal}}\)
on the balance only. Notice that baseline covariates do \textbf{not}
enter the magnitude: they affect the \emph{balance} between contraction
and growth but not the overall speed of the dynamics.

This restricted covariate structure --- covariates on the balance and on
the initial split, not on the magnitude --- is a deliberate
identification choice for the case study. With covariates entering all
three of
\(\eta^{\mathrm{tot}}_i, \eta^{\mathrm{bal}}_i, \eta^{\mathrm{init}}_i\)
unrestricted, the resulting three coefficient vectors are poorly
separated when SLD trajectories are short or noisy: a patient who
shrinks quickly looks the same as a patient with a fast magnitude and a
balance pushed entirely toward decrease. Routing covariates through
balance and initial split, while leaving magnitude to a population mean
plus arm and patient deviations, has empirically been the most
identifiable parameterisation across the trials we have fitted.

The covariate vector \(\mathbf{x}_i\) in
(Equation~\ref{eq-init-linpred}) and (Equation~\ref{eq-rate-linpred}) is
the same throughout. In the SCLC case study, \(\mathbf{x}_i\) comprises
six baseline covariates: age (years), sex (male indicator), haemoglobin
(g/dL), log-LDH, albumin (g/L), and ECOG performance status (\(\geq 1\)
indicator). We enter them through QR-rotated design matrices, with
weakly informative priors mapped through the QR rotation.

\subsubsection{Observation model}\label{sec-mech-observation}

At each post-baseline visit week \(t \in \mathcal{V}_i\) with
\(t > t_i^{\mathrm{bl}}\), we use a Gaussian observation density on the
log scale:
\begin{equation}\protect\phantomsection\label{eq-obs-density}{
\log\!\frac{y_i(t)}{y_i^{\mathrm{bl}}} \;\sim\; \mathtt{Normal}\!\big(\log B_i(t), \; \sigma_y^2\big),
}\end{equation} with observation scale \(\sigma_y\) shared across
patients and visits. The baseline visit itself is excluded from the
likelihood: because we condition on \(y_i^{\mathrm{bl}}\) as the
normaliser in (Equation~\ref{eq-sld-normalised}), the baseline
observation defines the patient's reference scale and is not modelled as
a noisy outcome of the dynamics. The log-scale Gaussian assumption is
consistent with the multiplicative measurement-error structure of RECIST
1.1 --- multiple target lesions, inter-reader variability, and
discreteness of the sum-of-diameters all contribute proportionally to
the size of the lesions being measured --- and yields tractable
left-censoring under the limit of detection (below). A heavier-tailed
alternative (Student-\(t\)) would be a natural robustness extension; we
discuss this briefly in Section~\ref{sec-discussion}.

When SLD falls below the assay limit of detection (in practice, when all
target lesions disappear so that \(y_i(t) = 0\) would be reported), the
observation is treated as left-censored at the limit-of-detection value
\(y^{\mathrm{LoD}}\), with the corresponding contribution to the
likelihood being the Gaussian CDF at
\(\log(y^{\mathrm{LoD}} / y_i^{\mathrm{bl}})\) rather than the density.
This handling of LoD is essential in IO trials where deep responses are
common; treating the absence as a measurement of zero or omitting the
visit week both introduce bias.

\subsubsection{Outputs that flow to the
multistate}\label{sec-mech-outputs}

For each posterior draw, the mechanistic submodel delivers for every
patient and every week: the latent burden \(B_i(t)\), and the change
rates \(r^{\mathrm{dec}}_i, r^{\mathrm{gro}}_i\). These are the inputs
to the time-varying covariate bridge \(\mathbf{W}_i(t)\) defined in
Section~\ref{sec-multistate}.

\subsection{Multistate proportional-hazard
submodel}\label{sec-multistate}

\subsubsection{State graph and time scales}\label{sec-ms-states}

The multistate submodel governs the stochastic clinical events. Each
patient occupies one of four states at every week \(t\): alive and
progression-free (state \(0\)), progressed and alive (state \(1\)), dead
(state \(2\), absorbing), or off-trial without observed death (state
\(3\)); the five admissible transitions between these states, together
with their time-scale conventions, are shown in
Figure~\ref{fig-multistate-diagram} below. Each transition has its own
intensity, with its own time scale and its own dependence on covariates
and on the latent burden trajectory.

\begin{figure}

\centering{

\includegraphics[width=0.85\linewidth,height=\textheight,keepaspectratio]{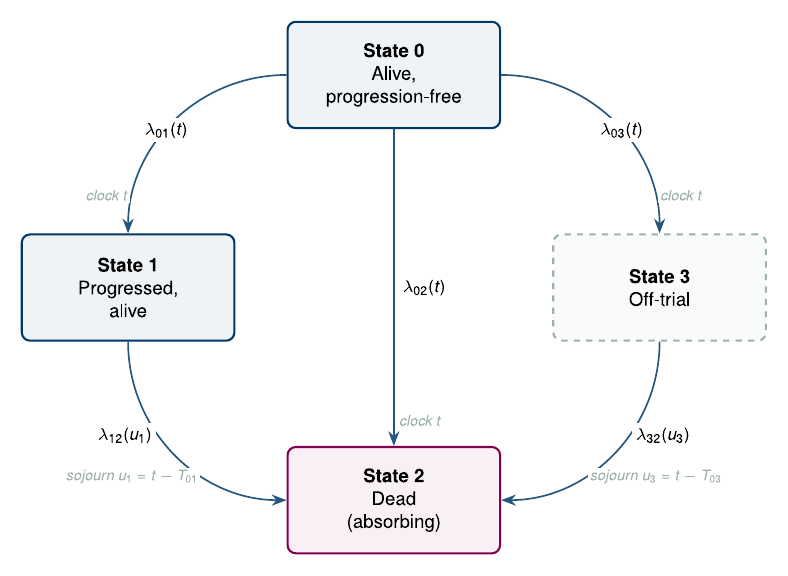}

}

\caption{\label{fig-multistate-diagram}\textbf{Multistate state graph.}
The five transition intensities used in the case study. Three
transitions out of state \(0\) (\(0\to 1\) progression, \(0\to 2\)
direct death, \(0\to 3\) dropout) use clock-forward time \(t\). The two
transitions into state \(2\) from non-zero source states (\(1\to 2\)
post-progression death, \(3\to 2\) off-trial death) use sojourn time
\(u_1 = t - T_{01}\) and \(u_3 = t - T_{03}\) respectively (semi-Markov;
the clock resets on entry into the source state). State \(3\) is dashed
to mark its off-trial status; state \(2\) is filled to mark it as the
absorbing terminal state.}

\end{figure}%

The case-study configuration has all five transitions active. The two
sojourn-clock transitions (\(1\to 2\) and \(3\to 2\)) use semi-Markov
time scales: the post-progression death hazard depends on time since
progression, not on calendar time; the off-trial death hazard depends on
time since dropout. Markov and additive-extended alternatives are
supported by the framework; we comment on the trade-offs at the end of
this subsection.

\subsubsection{Per-transition hazard}\label{sec-ms-hazard}

Write \(\lambda_{jk, i}(t)\) for patient \(i\)'s instantaneous
transition rate from state \(j\) to state \(k\) at \emph{clock time}
\(t\) (for clock-forward transitions) or at \emph{sojourn time}
\(t - T_{j,i}\) (for semi-Markov transitions, where \(T_{j,i}\) is the
entry time into state \(j\)). All five hazards share the same
proportional-hazards skeleton:
\begin{equation}\protect\phantomsection\label{eq-ms-hazard}{
\log \lambda_{jk, i}(t)
\;=\;
\underbrace{\log \lambda^{\mathrm{pop}}_{jk}(t)}_{\text{population baseline}}
\;+\;
\underbrace{\log \lambda^{\mathrm{trial}}_{jk, s[i]}(t)}_{\text{trial-level deviation}}
\;+\;
\underbrace{\boldsymbol{\beta}^{\mathrm{tv}}_{jk} \cdot \mathbf{W}_i(t)}_{\text{time-varying tumour bridge}}
\;+\;
\underbrace{\boldsymbol{\beta}^{\mathrm{ti}}_{jk} \cdot \mathbf{x}_i}_{\text{baseline covariates}}
\;+\;
\underbrace{\gamma_{jk,i}}_{\text{frailty}},
}\end{equation} in which only the first three terms are functions of
time. The population and trial-level baselines are Gaussian processes
(Section~\ref{sec-ms-gp}); the tumour bridge \(\mathbf{W}_i(t)\) is the
explicit coupling to the mechanistic submodel
(Section~\ref{sec-ms-bridge}); the baseline covariates \(\mathbf{x}_i\)
are the same patient-level covariate vector used in the mechanistic
submodel (Section~\ref{sec-mechanistic}), entered via QR-rotated design
matrices with weakly informative priors; and \(\gamma_{jk,i}\) is a
patient-level frailty term, described below.

Each of the five additive components is \emph{transition-specific}:
there is no constraint that, say, the coefficient on log-burden in the
\(0\to 1\) hazard equals the corresponding coefficient in the \(1\to 2\)
hazard. This is the structural point that distinguishes our framework
from a single-hazard joint model: the tumour-derived covariates are
allowed (under the priors) to act differently on progression hazard than
on post-progression death hazard, and the data decide which transitions
they actually inform.

It helps to fix the discrete-time reading of
(Equation~\ref{eq-ms-hazard}) before turning to a subtlety specific to
the tumour application. Because events are recorded on a weekly grid
rather than continuously, the natural object is the discrete-time
(grouped-duration) hazard familiar from econometric duration analysis
(Wooldridge 2010, ch.~22). Writing \(J_i(t) \in \{0, 1, 2, 3\}\) for the
state patient \(i\) occupies at week \(t\), the single-week conditional
survival is \(\exp(-\lambda_{jk,i}(t))\) (the \(b=a=t\) case of
\(S_{jk,i}\) in Section~\ref{sec-ms-likelihood}), so
\(\lambda_{jk,i}(t)\) is the \emph{cause-specific} rate of the
\(j\to k\) transition for a patient in source state \(j\), and the
per-week transition probability is its complement:
\begin{equation}\protect\phantomsection\label{eq-ms-discrete}{
\Pr\!\big[\, J_i(t) = k \;\big|\; J_i(t-1) = j,\; \mathbf{x}_i,\; \mathbf{W}_i(t) \big]
\;=\; 1 - \exp\!\big(-\lambda_{jk, i}(t)\big).
}\end{equation}

The subtlety is what the \(0\to 1\) hazard represents. RECIST 1.1
defines progression of \emph{target} lesions as a \(\geq 20\%\) increase
in SLD from nadir --- a criterion that is a deterministic function of
the latent burden trajectory \(B_i(t)\) and so requires no stochastic
hazard: at each posterior draw we know exactly when, if ever, the
modelled trajectory crosses the threshold. The mechanistic submodel
resolves this \emph{target-lesion} progression channel directly. What is
left for \(\lambda_{01,i}\) to model is the complementary
\emph{non-target} channel --- unequivocal progression of non-target
disease, the appearance of new lesions, and clinical deterioration ---
none of which is a function of the SLD trajectory. Let
\(T^{\mathrm{tgt}}_i = \min\{t : B_i(t) \ge 1.2\,\min_{u<t} B_i(u)\}\)
be the (per-draw, deterministic) week at which the latent trajectory
first meets the RECIST target-progression criterion. So
\(\lambda_{01,i}(t)\) is properly read as the rate of \emph{non-target}
progression among patients who have not already progressed on target
lesions, not as an undifferentiated progression hazard. The two channels
partition the routes to progression rather than competing to explain the
same event; the likelihood mechanics that enforce this are given in
Section~\ref{sec-ms-likelihood}.

One further feature distinguishes this transition from the others:
unlike death and dropout, which can occur --- and be recorded --- on any
week, non-target progression is \emph{detected} only at scheduled
assessment visits. Between assessments the patient is simply unobserved
for this channel, so no progression can be declared there even if the
underlying disease has progressed. We encode this by \textbf{gating the
hazard}, rather than by adding any new likelihood machinery. Using the
patient's scheduled-assessment week set \(\mathcal{V}_i\) (defined in
Section~\ref{sec-notation}), define the fixed (data, not estimated)
weekly indicator \(g_i(t) = \mathbf{1}[t \in \mathcal{V}_i]\). The
\emph{gated} non-target progression hazard is
\begin{equation}\protect\phantomsection\label{eq-ms-gate}{
\tilde{\lambda}_{01, i}(t) \;=\; g_i(t)\,\lambda_{01, i}(t),
}\end{equation} i.e.~the ordinary log-linear hazard of
(Equation~\ref{eq-ms-hazard}) on visit weeks and identically zero
between them. Substituting \(\tilde{\lambda}_{01,i}\) for
\(\lambda_{01,i}\) turns the generic per-week transition probability
(Equation~\ref{eq-ms-discrete}) into
\begin{equation}\protect\phantomsection\label{eq-ms-01-cond}{
\Pr\!\big[\, J_i(t) = 1 \;\big|\; J_i(t-1) = 0,\; T^{\mathrm{tgt}}_i > t,\; \mathbf{x}_i,\; \mathbf{W}_i(t) \big]
\;=\; 1 - \exp\!\big(-\tilde{\lambda}_{01, i}(t)\big),
}\end{equation} which is exactly \(1 - \exp(-\lambda_{01,i}(t))\) on
visit weeks and \(0\) between them --- progression cannot be detected
off-schedule. The gate multiplies the \emph{rate}; it is not a covariate
in the log-linear predictor, where a finite coefficient would merely
rescale the off-visit hazard rather than switch it off. With this one
substitution, the \(0\to 1\) channel reuses the \emph{identical}
discrete-time competing-risks likelihood as every other transition
(Section~\ref{sec-ms-likelihood}); the only difference is that its rate
vanishes off the assessment grid.

In the case-study configuration, two noteworthy choices are made
regarding (Equation~\ref{eq-ms-hazard}):

\begin{itemize}
\tightlist
\item
  \textbf{Dropout (\(0\to 3\)) carries both bridge and baseline
  covariates}, so both \(\boldsymbol{\beta}^{\mathrm{tv}}_{03}\) and
  \(\boldsymbol{\beta}^{\mathrm{ti}}_{03}\) are estimated. The TV
  component lets dropout risk track the latent tumour trajectory
  (patients on a deteriorating trajectory may withdraw faster); the TI
  component captures static baseline risk factors. Priors on
  \(\boldsymbol{\beta}^{\mathrm{tv}}_{03}\) are kept tight given the
  modest number of dropout events.
\item
  \textbf{Post-dropout death (\(3\to 2\)) has no covariate effects}, so
  \(\boldsymbol{\beta}^{\mathrm{tv}}_{32} = \boldsymbol{\beta}^{\mathrm{ti}}_{32} = \mathbf{0}\).
  Off-trial follow-up is sparse and the competing hazard structure is
  already modelled via the sojourn-clock GP; adding covariate effects
  would not be identifiable.
\end{itemize}

The frailty terms \(\gamma_{jk,i}\) capture residual patient-level
heterogeneity in transition risk that is not explained by the tumour
bridge or the observed baseline covariates. In the case-study
configuration, frailty terms are enabled for the \(0\to 1\) and
\(0\to 3\) transitions at the patient level; the terms for the remaining
transitions are fixed at zero. Crucially, the two active frailty terms
are modelled \emph{jointly}: letting
\(\boldsymbol{\gamma}_i = (\gamma_{01,i},\, \gamma_{03,i})^\top\), we
place \begin{equation}\protect\phantomsection\label{eq-ms-frailty}{
\boldsymbol{\gamma}_i \;\sim\; \mathtt{Normal}\!\big(\mathbf{0},\; \boldsymbol{\Sigma}_\gamma\big),
\qquad
\boldsymbol{\Sigma}_\gamma = \mathrm{diag}(\boldsymbol{\sigma}_\gamma)\, \mathbf{R}_\gamma\, \mathrm{diag}(\boldsymbol{\sigma}_\gamma),
}\end{equation} where
\(\boldsymbol{\sigma}_\gamma = (\sigma_{01}, \sigma_{03})^\top\) are
per-transition marginal standard deviations and \(\mathbf{R}_\gamma\) is
a \(2\times 2\) correlation matrix with an LKJ prior. The off-diagonal
element of \(\mathbf{R}_\gamma\) captures whether patients who are
intrinsically high-risk for progression tend also to be high-risk for
dropout --- a cross-transition dependence that neither the tumour bridge
nor the observed covariates can absorb. Sampling is performed via the
NCP Cholesky parameterisation of \(\boldsymbol{\Sigma}_\gamma\).

\subsubsection{Tumour-to-hazard bridge}\label{sec-ms-bridge}

The bridge \(\mathbf{W}_i(t)\) is the time-varying covariate vector
through which the mechanistic submodel feeds the multistate. In the
case-study configuration it has three components, all derived
deterministically from the mechanistic submodel's outputs:
\begin{equation}\protect\phantomsection\label{eq-ms-bridge}{
\mathbf{W}_i(t)
\;=\;
\bigg(
\underbrace{\tilde{B}_i(t)}_{\text{standardised log-burden}},\;
\underbrace{\log r^{\mathrm{dec}}_i}_{\text{log decrease rate}},\;
\underbrace{\log r^{\mathrm{gro}}_i}_{\text{log growth rate}}
\bigg)^{\!\top},
}\end{equation} where the standardised log-burden is
\begin{equation}\protect\phantomsection\label{eq-ms-bridge-sld}{
\tilde{B}_i(t) \;=\; \frac{\log B_i(t) + \log y_i^{\mathrm{bl}} - m^{\mathrm{sld}}}{q^{\mathrm{sld}}},
}\end{equation} with \(m^{\mathrm{sld}}\) and \(q^{\mathrm{sld}}\) the
median and inter-quartile range of \(\log y_i(t)\) across all observed
post-baseline visit weeks \(t \in \mathcal{V}_i\), computed once on the
data and held fixed during inference. The \(\log y_i^{\mathrm{bl}}\)
term in (Equation~\ref{eq-ms-bridge-sld}) re-attaches the patient's
baseline scale, so \(\tilde{B}_i(t)\) is informative about the
\emph{absolute} tumour size at week \(t\) rather than only its
fold-change relative to baseline.

Two design choices in (Equation~\ref{eq-ms-bridge}) are worth flagging.

First, the bridge uses the \emph{latent} burden \(B_i(t)\) from
(Equation~\ref{eq-sld-decomp}), not the observed SLD trajectory. The
latent burden is smooth, defined at every week, and free of measurement
noise; the observed trajectory is noisy and defined only at scheduled
visits. Using the latent quantity as the time-varying covariate is what
makes the joint model \emph{joint}: the multistate likelihood at week
\(t\) depends on the posterior over \(B_i(t)\), so progression times and
death times inform the latent state's posterior, and vice-versa.

Second, the \emph{baseline} rates \(\log r^{\mathrm{dec}}_i\) and
\(\log r^{\mathrm{gro}}_i\) are patient-specific but time-invariant; the
Gompertz decay mechanism in Section~\ref{sec-mech-gompertz} makes the
\emph{effective} growth rate time-varying at the trajectory level, but
the bridge receives the latent burden \(B_i(t)\) directly, which already
incorporates that attenuation. The rates appear in \(\mathbf{W}_i(t)\)
rather than in \(\mathbf{x}_i\) because the framework also supports the
AR(1) variant in which they become explicitly time-varying (cf.~the
footnote in Section~\ref{sec-mech-dynamics}); the bridge is structured
to accommodate that extension without changing the multistate submodel's
interface.

In the case-study configuration, the bridge feeds the \(0\to 1\)
progression and \(0\to 3\) dropout hazards. The framework also supports
feeding it into the direct-death (\(0\to 2\)) and post-progression death
(\(1\to 2\)) hazards; we do not exercise these in the case study for the
identification reasons given in Section~\ref{sec-ms-hazard}. The
resulting bridge coefficients are reported in the bottom panel of
Figure~\ref{fig-covariate-effects}.

\subsubsection{Gaussian-process baselines}\label{sec-ms-gp}

The two time-dependent log-baselines in (Equation~\ref{eq-ms-hazard})
are modelled as Gaussian processes. For each transition \(jk\) the
\emph{population baseline} is \[
\log \lambda^{\mathrm{pop}}_{jk}(\cdot) \;\sim\; \mathtt{GP}\!\big(\mu^{\mathrm{pop}}_{jk}, \;k_{jk}(\cdot, \cdot;\, \alpha^{\mathrm{pop}}_{jk},\, \rho^{\mathrm{pop}}_{jk})\big),
\] with squared-exponential covariance kernel \[
k_{jk}(t, t';\, \alpha, \rho) \;=\; \alpha^2 \exp\!\bigg(\!-\frac{(t - t')^2}{2 \rho^2}\bigg),
\] intercept \(\mu^{\mathrm{pop}}_{jk}\) (population log-baseline-hazard
level), marginal SD \(\alpha^{\mathrm{pop}}_{jk}\), and length scale
\(\rho^{\mathrm{pop}}_{jk}\). The trial-level baseline
\(\log \lambda^{\mathrm{trial}}_{jk, s[i]}(\cdot)\) is an analogous GP
with its own marginal SD and length scale, mean zero, and one
independent realisation per trial. Patient-level GP residuals are not
used in the case study.

For the three clock-forward transitions (\(0\to 1\), \(0\to 2\),
\(0\to 3\)), the GP is defined on calendar time
\(t \in \{1, 2, \dots, T\}\) with \(T\) the maximum study week across
all patients. For the two sojourn-clock transitions (\(1\to 2\),
\(3\to 2\)), the GP is defined on sojourn time
\(u \in \{1, 2, \dots, U_{jk}\}\) with \(U_{jk}\) the maximum sojourn
time observed (or required for forecasting). The two clocks have
separate GPs with independent hyperparameters; the GP defined on \(t\)
is shared across patients (same calendar time means the same
baseline-hazard value), and the GP defined on \(u\) is shared across
patients in the same source state (a patient who has been in state \(1\)
for \(u\) weeks has the same post-progression death baseline hazard as
any other such patient, regardless of when their progression occurred).

For computational efficiency the GPs are evaluated on a coarse knot grid
of step size \(\Delta_{\mathrm{GP}}\) weeks (eight in the case study)
and pushed to weekly resolution by piecewise-constant assignment. The GP
draws are mean-centred so that the intercept \(\mu^{\mathrm{pop}}_{jk}\)
alone controls the level; without this centring, the intercept and the
GP's overall offset are not separately identifiable.

\subsubsection{Likelihood}\label{sec-ms-likelihood}

For each patient \(i\), the multistate submodel contributes a likelihood
term that is a product of survival probabilities and instantaneous
hazards along the patient's observed trajectory through the state graph.
Writing
\(S_{jk, i}(a, b) = \exp\!\big(\!-\sum_{u = a}^{b} \lambda_{jk, i}(u)\big)\)
for patient \(i\)'s discrete-time conditional survival across the
interval \([a, b]\), the contribution depends on the patient's
\emph{terminal pattern} observed by the data cut-off:

\begin{itemize}
\tightlist
\item
  \emph{Censored in state \(0\):} \(S_{01,i} S_{02,i} S_{03,i}\) over
  the patient's full at-risk window in state \(0\).
\item
  \emph{Progressed, alive at cut-off (state \(1\)):} survival in state
  \(0\) until detection of progression, instantaneous hazard
  \(\lambda_{01,i}\) at the progression week, then survival under
  \(\lambda_{12,i}\) over the observed sojourn in state \(1\).
\item
  \emph{Died after progression (state \(2\) via \(1\)):} as above, plus
  \(\lambda_{12,i}\) at the death week.
\item
  \emph{Died without progression (state \(2\) via \(0\to 2\)):} survival
  in state \(0\) until death, with \(\lambda_{02,i}\) at the death week.
\item
  \emph{Off-trial without death (state \(3\)):} survival in state \(0\)
  until dropout, \(\lambda_{03,i}\) at the dropout week, then survival
  under \(\lambda_{32,i}\) over the observed sojourn in state \(3\).
\item
  \emph{Died off-trial (state \(2\) via \(3\)):} as above, plus
  \(\lambda_{32,i}\) at the death week.
\end{itemize}

The classification of each patient into one of these terminal patterns
is performed deterministically from the trial data and does not involve
any model parameters; the patterns are stored alongside the transition
times in the data passed to Stan. Within a pattern, the only
model-dependent quantities are the conditional survivals and the hazards
at observed event times, which are computed directly from the
per-transition log-hazards (Equation~\ref{eq-ms-hazard}).

Two features of the case-study likelihood deserve a brief note.

\begin{figure}

\centering{

\includegraphics[width=0.95\linewidth,height=\textheight,keepaspectratio]{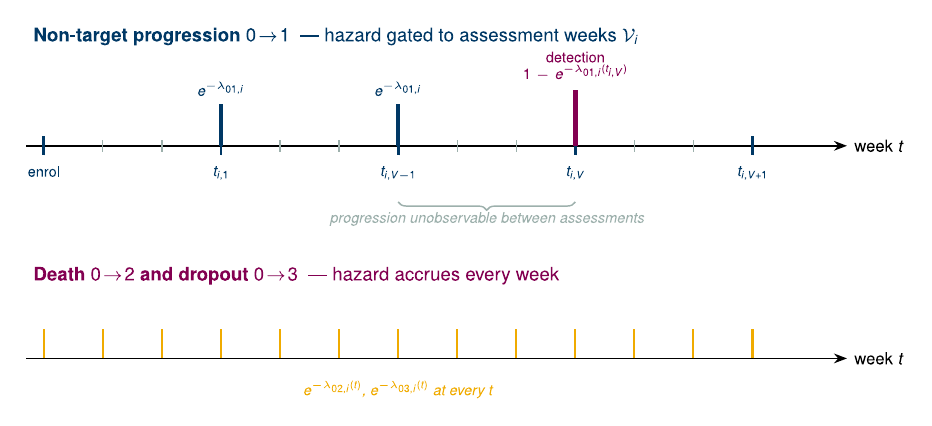}

}

\caption{\label{fig-visit-gating}\textbf{Visit-gating of the non-target
progression channel.} \textbf{Top:} the \(0\to 1\) hazard
\(\lambda_{01,i}\) contributes a single-assessment survival factor
\(e^{-\lambda_{01,i}}\) only at the patient's scheduled assessment weeks
\(\mathcal{V}_i\) (bold navy ticks), and a detection event
\(1 - e^{-\lambda_{01,i}}\) can land only at such a visit (plum). The
weeks between assessments (faint grey) carry no progression hazard ---
progression that occurs there cannot be observed until the next visit,
so the true event time is known only up to the inter-assessment gap.
\textbf{Bottom:} death (\(0\to 2\)) and dropout (\(0\to 3\)) are
observed on whichever week they occur, so their hazards accrue on every
week of the grid. The two channels therefore tick on different clocks:
progression on the coarse assessment grid, death and dropout on the full
weekly grid.}

\end{figure}

\emph{Visit-gating of non-target progression.} The \(0\to 1\) channel
uses the same survival-times-hazard contribution as every other
transition; the only change is that it runs on the gated hazard
\(\tilde{\lambda}_{01,i}(t) = g_i(t)\,\lambda_{01,i}(t)\) of
(Equation~\ref{eq-ms-gate}) rather than on \(\lambda_{01,i}\) itself.
Because the gate sets the rate to zero between assessments, this
substitution makes the generic likelihood collapse to a product over
visit weeks with no special-casing. For a patient whose non-target
progression is confirmed at scheduled visit \(V\), write
\(S_{01,i}(1, t_{i,V}-1) = \exp(-\sum_{u=1}^{t_{i,V}-1} \tilde{\lambda}_{01,i}(u))\)
as in Section~\ref{sec-ms-likelihood}; every off-visit week contributes
a zero exponent, so the survival sum telescopes onto the assessment grid
and the contribution is
\begin{equation}\protect\phantomsection\label{eq-ms-01-visit-gated}{
\underbrace{\prod_{v\,:\, t_{i,v} < t_{i,V}} \exp\!\big(-\lambda_{01,i}(t_{i,v})\big)}_{\text{survived every prior assessment}}
\;\times\;
\underbrace{\Big(1 - \exp\!\big(-\lambda_{01,i}(t_{i,V})\big)\Big)}_{\text{detected at visit } V}.
}\end{equation} The product runs over the patient's \emph{visit weeks}
strictly before detection because the intervening weeks have
\(\tilde{\lambda}_{01,i} = 0\) and hence survival factor \(1\); the
detection term is the ordinary complement
(Equation~\ref{eq-ms-discrete}) evaluated at the visit week \(t_{i,V}\),
where \(g_i = 1\). A patient still progression-free in state \(0\) at
the data cut-off contributes only the survival product, over all
assessment weeks up to the cut-off. The weekly grid still carries the
patient's risk of death and dropout; it is only the \emph{progression}
channel whose rate vanishes off the assessment grid.
Figure~\ref{fig-visit-gating} illustrates the construction against the
continuously-observed death and dropout channels.

Two practical points complete the picture. First, the covariates feeding
\(\lambda_{01,i}\) --- log burden and the decrease/growth velocities
carried by the bridge \(\mathbf{W}_i(t)\) (Section~\ref{sec-ms-bridge})
--- are read from the \emph{latent} modelled trajectory \(B_i(t)\) at
each visit week, not from the noisy observed SLD. Besides the
joint-modelling rationale of Section~\ref{sec-ms-bridge}, this is also
what makes the visit-gated hazard numerically well-behaved: at a
complete-response visit the observed SLD is zero and
\(\log(\text{SLD})\) is undefined, whereas the latent \(B_i(t)\) is
bounded away from zero by the limit-of-detection floor. Second,
target-lesion progression is \emph{not} gated this way: it is read off
the latent trajectory deterministically (next paragraph) and carries no
\(\lambda_{01,i}\) term at all.\footnote{Gating makes \(\lambda_{01,i}\)
  an assessment-\emph{anchored} detection rate rather than a
  continuous-time biological hazard, so its scale depends on how often a
  patient is assessed --- the same underlying disease accrues more
  per-visit hazard under a six-weekly schedule than under a weekly one.
  We absorb the resulting cross-trial confound through the hierarchy
  rather than the likelihood: the \(0\to 1\) baseline hazard carries a
  trial-arm--level random intercept \emph{and} its own Gaussian-process
  residual (Section~\ref{sec-ms-gp}), so a constant cadence-induced
  level shift is soaked up at the trial-arm level rather than mistaken
  for faster progression. Two residual effects remain --- the
  coefficient magnitudes in \(\lambda_{01,i}\) are interpretable only
  relative to the assessment grid, not as per-week rates, and
  unconditional (\(\mathrm{spop}\)) forecasts regenerate progression on
  a fixed six-weekly grid (matching the case-study cadence), imposing a
  granularity of up to one assessment interval on predicted progression
  times. Both are properties of interval-censored progression data, not
  of the model, and neither affects the death or dropout channels, which
  remain on the weekly grid.}

\emph{Target-lesion progression handled deterministically.} As noted in
Section~\ref{sec-ms-hazard}, the RECIST \(\geq 20\%\)-from-nadir
criterion for target lesions is a deterministic function of the latent
burden trajectory \(B_i(t)\) at each posterior draw
(Section~\ref{sec-mechanistic}). Patients whose trajectory crosses this
threshold within their observed follow-up window have their \(0\to 1\)
transition resolved on the mechanistic side, not by \(\lambda_{01,i}\);
the data flag the week and the likelihood applies \emph{no}
\(\lambda_{01,i}\) hazard factor at it (these patients still contribute
survival under \(\lambda_{01,i}\) up to that week, since they remained
at risk of non-target progression until then). This is what makes
\(\lambda_{01,i}\) the \emph{non-target}, target-progression-free hazard
of Section~\ref{sec-ms-hazard} rather than an all-cause progression
hazard, and it avoids double-counting patients whose progression the
mechanistic side already explains.

\subsubsection{Posterior-predictive endpoint
derivation}\label{sec-ms-forecast}

Given a single posterior draw of all model parameters, clinical
endpoints for patient \(i\) are obtained by forward-simulating the joint
model from the patient's last observed state. The latent burden
trajectory \(B_i(t)\) is extended using the mechanistic state-space
equations (Section~\ref{sec-mechanistic}), and the per-transition
hazards \(\lambda_{jk,i}(t)\) are computed from
(Equation~\ref{eq-ms-hazard}) using the extended bridge
\(\mathbf{W}_i(t)\). This is the \emph{conditional} forecasting mode
(denoted \(\mathrm{sample}\)): for each patient, the model retains their
observed event history through the data cut-off --- who has already
progressed, died, or dropped out --- and forward-simulates only the
unresolved future. Patients who already experienced an event contribute
their actual event time directly; only those still censored at the
cut-off enter the routing tree below, from the cut-off forward. This is
the operationally relevant mode: it answers the question a trialist
faces at an interim look --- ``given everything we know about these
patients today, what will the mature survival curve look like?'' --- and
is the mode used throughout the leave-future-out evaluation
(Section~\ref{sec-results-lfo}).

As a model-checking diagnostic, the same routing can also be run
\emph{unconditionally} from treatment start (the \(\mathrm{spop}\)
estimand), discarding all observed event histories and re-simulating
every patient from week 1. This tests whether the model's generative
mechanism can reproduce the observed KM from inferred population
parameters alone --- a goodness-of-fit check rather than a forecasting
tool (Section~\ref{sec-results-survival}).

The routing logic is identical for both modes and is described below;
the conditional pass enters the tree at the patient's last observed
state, while the unconditional pass enters from treatment start.

\textbf{Competing exits from state 0.} Progression, direct death, and
dropout are \emph{competing risks}: they share one risk set, and the
earliest sampled event determines how the patient leaves state \(0\). We
sample a candidate time for each:

\begin{itemize}
\tightlist
\item
  \textbf{Progression (\(0\to 1\))} at
  \(T^{01}_i = \min(T^{\mathrm{tgt}}_i,\, T^{\mathrm{ms}}_i)\), itself
  the earlier of two sub-channels: the deterministic target-lesion time
  \(T^{\mathrm{tgt}}_i\) (Section~\ref{sec-ms-hazard}, the first week
  the extended latent trajectory meets the RECIST threshold) and the
  stochastic non-target time \(T^{\mathrm{ms}}_i\) (sampled from the
  gated hazard \(\tilde{\lambda}_{01,i}\), which fires only at visit
  weeks \(t \in \mathcal{V}_i\), per Equation~\ref{eq-ms-gate}).
\item
  \textbf{Direct death (\(0\to 2\))} at \(T^{02}_i\), sampled from
  \(\lambda_{02,i}\).
\item
  \textbf{Dropout (\(0\to 3\))} at \(T^{03}_i\), sampled from
  \(\lambda_{03,i}\).
\end{itemize}

The realised exit is the transition with the smallest candidate time
(ties broken \(0\to 1 \succ 0\to 2 \succ 0\to 3\)). Crucially, a patient
may never progress: if the direct-death or dropout candidate is
earliest, the patient leaves state \(0\) without ever entering state
\(1\).

\textbf{Reading off PFS.} PFS is a progression-\emph{or}-death endpoint,
so it is determined by the exit route, not computed separately:
\begin{equation}\protect\phantomsection\label{eq-pfs-route}{
T^{\mathrm{pfs}}_i \;=\;
\begin{cases}
T^{01}_i & \text{progression wins } (0\to 1),\\
T^{02}_i & \text{direct death wins } (0\to 2),\\
\text{censored at } T^{03}_i & \text{dropout wins } (0\to 3).
\end{cases}
}\end{equation} Direct death without progression \emph{is} a PFS event
(PFS fails at the death time); dropout right-censors PFS at the
off-trial week.

\textbf{Reading off OS.} OS continues past the state-\(0\) exit along
the surviving path:

\begin{itemize}
\tightlist
\item
  \emph{Progression (\(0\to 1\)):} the patient enters state \(1\) at
  \(T^{01}_i\); OS is then a further draw from \(\lambda_{12,i}\), on
  the sojourn clock \(u_1 = t - T^{01}_i\) (semi-Markov) or calendar
  time \(t\) (Markov), depending on the configuration.
\item
  \emph{Direct death (\(0\to 2\)):} OS equals \(T^{02}_i\) --- the same
  event terminates both PFS and OS.
\item
  \emph{Dropout (\(0\to 3\)):} the patient enters state \(3\) at
  \(T^{03}_i\); off-trial death is a draw from \(\lambda_{32,i}\) on the
  sojourn clock \(u_3 = t - T^{03}_i\), otherwise OS is censored.
\end{itemize}

\textbf{Reading off response (ORR).} Objective response is not a
time-to-event read off the routing above; it is a best-response
\emph{classification} derived from the \emph{shape} of the latent burden
trajectory, and so comes entirely from the mechanistic submodel rather
than the multistate hazards. For each draw we evaluate the extended
trajectory \(B_i(t)\) over the patient's on-trial window and apply the
RECIST 1.1 response rules to the fold-change from baseline, \(B_i(t)\)
(recall \(B_i\) is normalised so \(B_i(t_i^{\mathrm{bl}}) = 1\)): a
complete response (CR) when the trajectory reaches the
limit-of-detection floor, a partial response (PR) when it falls at least
\(30\%\) below baseline, progressive disease (PD) at the target-lesion
threshold \(T^{\mathrm{tgt}}_i\) already defined above, and stable
disease (SD) otherwise. The objective response indicator is
\(\mathbf{1}[\text{best response} \in \{\mathrm{CR}, \mathrm{PR}\}]\),
optionally subject to a confirmation rule (a qualifying response
sustained to the next scheduled assessment), and ORR is the cohort
average. The same latent \(B_i(t)\) thus produces both the progression
\emph{timing} that feeds the competing-risks routing and the response
\emph{depth} that feeds ORR --- they are two readings of one trajectory.

\textbf{Endpoints as posterior functionals.} Kaplan--Meier curves for
PFS and OS are computed from the
\(\{T^{\mathrm{pfs}}_i, T^{\mathrm{os}}_i\}\) sample paths, ORR from the
per-patient response classification, and any other endpoint analogously
--- each averaged across posterior draws. Because every quantity flows
from the same joint posterior, each reported endpoint inherits the
model's full parameter uncertainty; no re-fitting or plug-in step is
involved.\footnote{A Kaplan--Meier credible band and a scalar- or
  quantile-endpoint interval (e.g.~median PFS) are two readings of the
  \emph{same} posterior over sample paths, and they display its
  uncertainty along orthogonal axes. The KM band is read
  \textbf{vertically}: at each fixed time \(t\) it shows the credible
  range of the survival probability \(S(t)\). A median- (or any
  quantile-) PFS interval is read \textbf{horizontally}: at the fixed
  probability \(S = 0.5\) it shows the credible range of the \emph{time}
  at which the curve crosses that level. The two are linked by the
  curve's local slope --- a first-order Jacobian,
  \(\Delta t \approx \Delta S / |S'(t)|\) --- so where the survival
  curve is steep (the high-hazard early window typical of these cohorts)
  a wide vertical band maps to a narrow horizontal interval, and where
  it is flat the reverse holds. Consequently a tight median-PFS interval
  sitting beside a visually broader KM ribbon is not a contradiction or
  an artefact of overconfidence; it is the expected consequence of
  resolving one posterior along a steep portion of the curve. The same
  correspondence governs the fixed-timepoint PFS probabilities (vertical
  reading) versus the median and quantile summaries (horizontal reading)
  reported below.}

\clearpage\begin{landscape}

\begin{figure}

\centering{

\includegraphics[width=1\linewidth,height=\textheight,keepaspectratio]{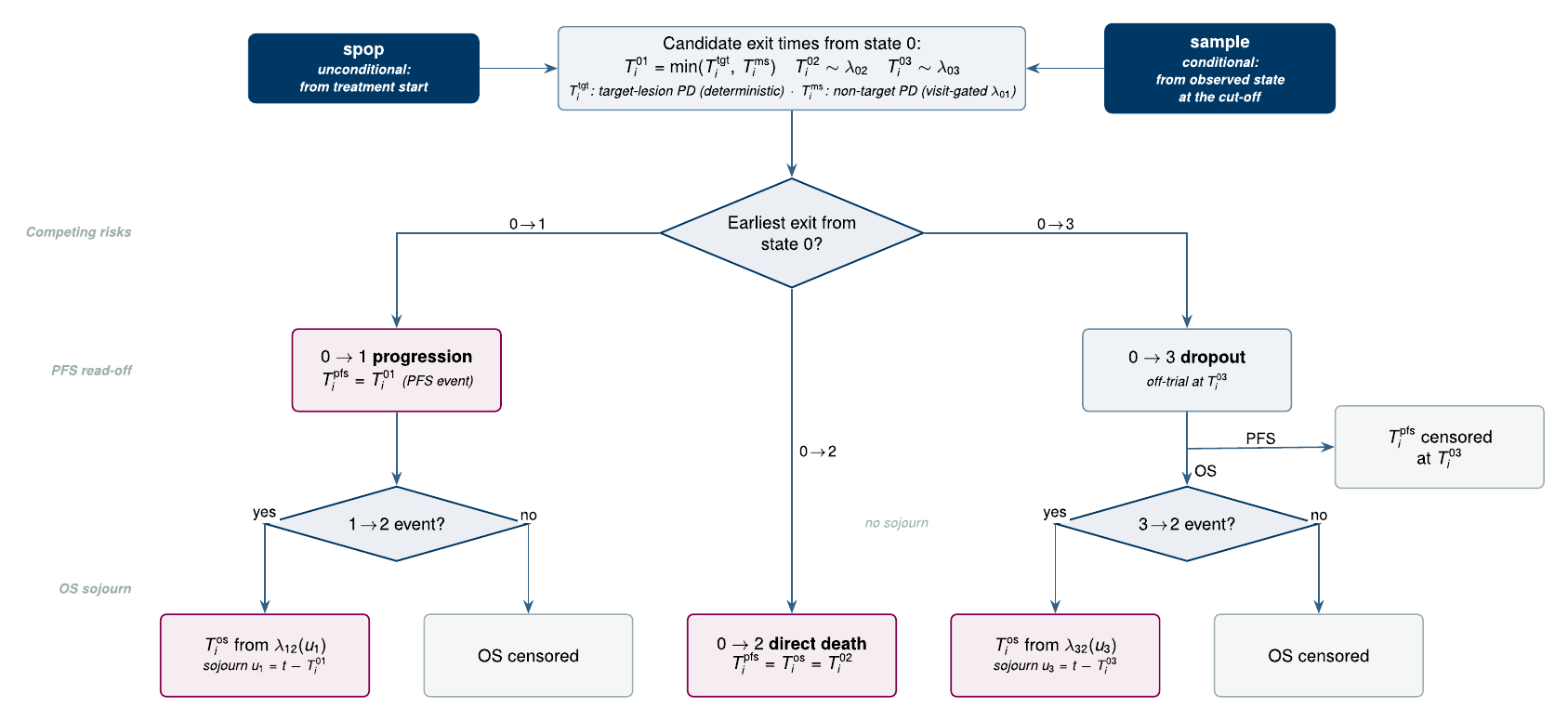}

}

\caption{\label{fig-gq-routing}\textbf{Posterior-predictive endpoint
routing for a single posterior draw.} The unconditional
(\(\mathrm{spop}\)) and conditional (\(\mathrm{sample}\)) passes feed
the \emph{same} routing tree and differ only in where they enter it ---
from treatment start, or from the observed state at the data cut-off.
The three competing exits from state 0 are resolved first; the earliest
candidate time wins. Progression's own candidate time is the minimum of
the deterministic target-lesion time \(T^{\mathrm{tgt}}_i\) and the
stochastic non-target time \(T^{\mathrm{ms}}_i\). PFS and OS are then
read off the realised route: direct death is itself a PFS event, dropout
right-censors PFS, and only the progression route incurs a
post-progression sojourn before death.}

\end{figure}%

\end{landscape}

\textbf{Figures.}

\begin{itemize}
\tightlist
\item
  Fig. 4.1 --- multistate state diagram (this section,
  Figure~\ref{fig-multistate-diagram}).
\item
  Fig. 4.2 --- illustrative posterior baseline hazards
  \(\lambda_{01}, \lambda_{02}, \lambda_{12}\) from SCLC-01.
\item
  The bridge \(\mathbf{W}_i(t)\) coefficient forest plot is shown with
  the results in the bottom panel of Figure~\ref{fig-covariate-effects}.
\end{itemize}

\section{Inference and implementation}\label{sec-inference}

Full Bayesian inference is performed via Hamiltonian Monte Carlo (HMC)
using the No-U-Turn Sampler (NUTS) as implemented in Stan 2.38, accessed
through the CmdStanR interface in R. Four parallel chains are run for
each model fit, with 150 warmup iterations (warm-started from previously
adapted inverse mass matrices) and 500 sampling iterations per chain,
yielding 2000 posterior draws. Convergence is assessed via
split-\(\hat{R}\) (target \(< 1.01\)), bulk and tail effective sample
size (ESS), and the absence of divergent transitions after warmup. The
non-centred parameterisation used throughout the hierarchy is essential
for efficient sampling of the weakly-identified patient-level
parameters. Computational scalability is achieved through three
strategies: the Gaussian process baselines are evaluated on a coarse
knot grid (8-week spacing, \textasciitilde25 knots per transition) and
interpolated to the weekly resolution required by the likelihood;
disabled transitions are sized to zero-length arrays so they incur no
computational cost; and the model is compiled with threading support for
within-chain parallelism. A single posterior fit for the SCLC case study
(\(N = 497\) patients, 5 active transitions, \textasciitilde200k
parameters) completes in approximately 3--4 hours on a 12-core machine.
The full analysis pipeline --- from raw data through model compilation,
MCMC sampling, and endpoint derivation --- is orchestrated via the R
\texttt{targets} package, ensuring reproducibility and caching of
intermediate results across iterative development cycles.

\section{Case study: data}\label{sec-data}

\subsection{Trials and disease
context}\label{trials-and-disease-context}

\subsubsection{Small-cell lung cancer}\label{sec-data-sclc}

We apply PIONEER to two completed Phase II/III trials in first-line
extensive-stage small-cell lung cancer (ES-SCLC; referred to throughout
as \textbf{SCLC} in text, tables, and figures), both employing
platinum-etoposide backbone chemotherapy:

\begin{itemize}
\tightlist
\item
  \textbf{Lilly CXCR4} (Study I2V-MC-CXAC): a randomised Phase II trial
  of the CXCR4 inhibitor LY2510924 plus carboplatin/etoposide (Arm A,
  \(N=40\)) versus carboplatin/etoposide alone (Arm B, \(N=37\)). This
  serves as the \emph{target trial} whose endpoints we aim to forecast.
\item
  \textbf{Amgen 20010145}: a randomised Phase III trial of darbepoetin
  alfa (NESP, \(N=214\)) versus placebo (\(N=205\)), both in combination
  with platinum/etoposide. Darbepoetin is an erythropoiesis-stimulating
  agent (for chemotherapy-induced anaemia) with no anticipated
  anti-tumour effect. This serves as the \emph{historical trial} from
  which the population-level tumour dynamics are borrowed.
\end{itemize}

Both trials share the same disease (ES-SCLC), the same standard-of-care
backbone (platinum/etoposide), and the same clinical trajectory (high
initial response rate followed by rapid relapse). This commonality
justifies the hierarchical pooling assumption: population-level tumour
growth kinetics are expected to be comparable across the two trials,
with trial-level random effects absorbing any residual heterogeneity.

\subsection{Data source}\label{data-source}

This publication is based on research using information obtained from
\href{http://www.projectdatasphere.org}{www.projectdatasphere.org},
which is maintained by Project Data Sphere. Neither Project Data Sphere
nor the owner(s) of any information from the web site have contributed
to, approved or are in any way responsible for the contents of this
publication.

\subsection{Data processing}\label{data-processing}

A standardized pipeline harmonized all trials into a common analytical
schema. For each trial, a dedicated extraction module accommodated the
sponsor-specific data format: SLD was extracted from tumour assessment
records at each scheduled visit, and patient-level summaries
(demographics, treatment arm, survival endpoints) were derived from the
clinical domains. The extracted data were mapped to two datasets per
trial:

\begin{enumerate}
\def\labelenumi{\arabic{enumi}.}
\tightlist
\item
  \textbf{Patient-level data}: one row per patient containing baseline
  covariates (age, sex, haemoglobin, log-LDH, albumin, ECOG performance
  status), treatment assignment, and survival outcomes (PFS and OS in
  weeks from treatment start; event indicators coded 1 = event, 0 =
  censored).
\item
  \textbf{Visit-level data}: one row per tumour assessment containing
  week from treatment start, SLD in millimetres, and the
  investigator-assessed RECIST 1.1 response (CR, PR, SD, PD, or NE).
\end{enumerate}

Missing baseline covariates were imputed at the population median prior
to model fitting. Patients were required to have at least one
post-baseline tumour assessment; those randomized but never treated or
lacking tumour measurements were excluded.

\subsection{Summary statistics}\label{summary-statistics}

\begin{table}

\caption{\label{tbl-trial-summary}Patient disposition and endpoints by
trial and arm --- SCLC.}

\centering{

\begin{tabular*}{1\linewidth}{@{\extracolsep{\fill}}l|rrrr}
\toprule
 & \multicolumn{2}{c}{Amgen 20010145} & \multicolumn{2}{c}{Lilly CXCR4} \\ 
\cmidrule(lr){2-3} \cmidrule(lr){4-5}
 & NESP + chemo & Placebo + chemo & Arm A (CXCR4 + chemo) & Arm B (chemo alone) \\ 
\midrule\addlinespace[2.5pt]
N & {214} & {205} & {41} & {37} \\ 
Median follow-up (mo) & {9.2} & {9.4} & {9.9} & {11.0} \\ 
Median PFS (mo) & {5.7} & {5.7} & {4.8} & {5.5} \\ 
PFS censored (\%) & {8\%} & {7\%} & {32\%} & {19\%} \\ 
Deaths & {170} & {174} & {30} & {24} \\ 
Visits per patient & {3.1} & {3.2} & {3.5} & {3.6} \\ 
\bottomrule
\end{tabular*}

}

\end{table}%

\begin{table}

\caption{\label{tbl-covariate-summary}Baseline covariate distributions
by trial --- SCLC.}

\centering{

\begin{tabular*}{\linewidth}{@{\extracolsep{\fill}}lcc}
\toprule
\textbf{Characteristic} & \textbf{Amgen 20010145}  N = 419\textsuperscript{\textit{1}} & \textbf{Lilly CXCR4}  N = 78\textsuperscript{\textit{1}} \\ 
\midrule\addlinespace[2.5pt]
Age (years) & 60.6 (8.8) & 65.3 (8.7) \\ 
Male (\%) & 278 (66\%) & 35 (45\%) \\ 
ECOG \ensuremath{\geq} 1 (\%) & 78 (19\%) & 4 (5.3\%) \\ 
    Unknown & 0 & 3 \\ 
Haemoglobin (g/dL) & 11.9 (1.0) & 13.2 (1.6) \\ 
    Unknown & 2 & 0 \\ 
log-LDH & 6.2 (0.7) & 5.8 (0.6) \\ 
    Unknown & 0 & 3 \\ 
Albumin (g/L) & 37.6 (4.8) & 35.1 (4.8) \\ 
    Unknown & 13 & 1 \\ 
\bottomrule
\end{tabular*}
\begin{minipage}{\linewidth}
\textsuperscript{\textit{1}}Mean (SD); n (\%)\\
\end{minipage}

}

\end{table}%

\begin{landscape}

\end{landscape}

\begin{figure}

\centering{

\pandocbounded{\includegraphics[keepaspectratio]{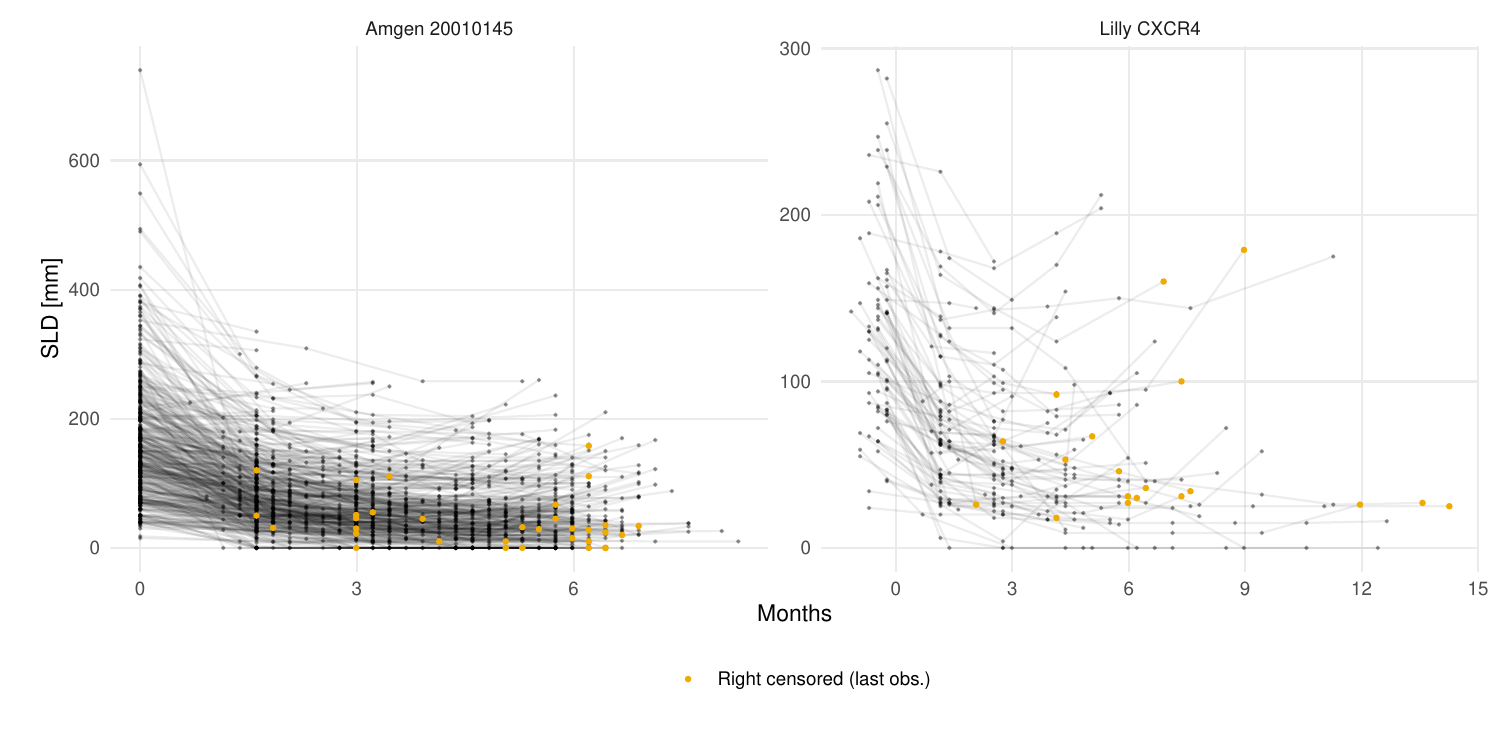}}

}

\caption{\label{fig-sld-trajectories-sclc}\textbf{SCLC: observed
tumour-burden trajectories.} Sum of target-lesion diameters (SLD) over
time for every patient, by trial. The last observation of each
right-censored patient (still progression-free at the data cut-off) is
highlighted; see legend.}

\end{figure}

The raw longitudinal tumour data motivate the mechanistic submodel
directly. Figure~\ref{fig-sld-trajectories-sclc} overlays every
patient's observed SLD trajectory, with the last observation of each
right-censored patient highlighted. Two features are immediately visible
and shape the model in Section~\ref{sec-mechanistic}: trajectories
follow a characteristic initial-decline-then-regrowth shape (the
empirical basis for the two-component \(D_i\)/\(G_i\) decomposition),
and a large fraction of patients are censored while their tumour is
still shrinking or near nadir --- these are precisely the immature
trajectories from which Section~\ref{sec-results-lfo} forecasts mature
survival.

\section{Results}\label{sec-results}

We present results from an application of PIONEER to a case study in
ES-SCLC (target trial: Lilly CXCR4, \(N = 78\); historical trial: Amgen
Darbe, \(N = 419\)). Results are organised around five evaluations:
model fit to the longitudinal tumour data
(Section~\ref{sec-results-fit}), forecasting of mature survival from
immature data via leave-future-out (LFO) validation
(Section~\ref{sec-results-lfo}), the clinical endpoint summaries that
this forecast implies (Section~\ref{sec-results-endpoints}),
unconditional posterior-predictive KM checks
(Section~\ref{sec-results-survival}), and transition-specific covariate
effects (Section~\ref{sec-results-covariates}).

\subsection{Model fit to longitudinal tumour
data}\label{sec-results-fit}

\begin{figure}

\centering{

\pandocbounded{\includegraphics[keepaspectratio]{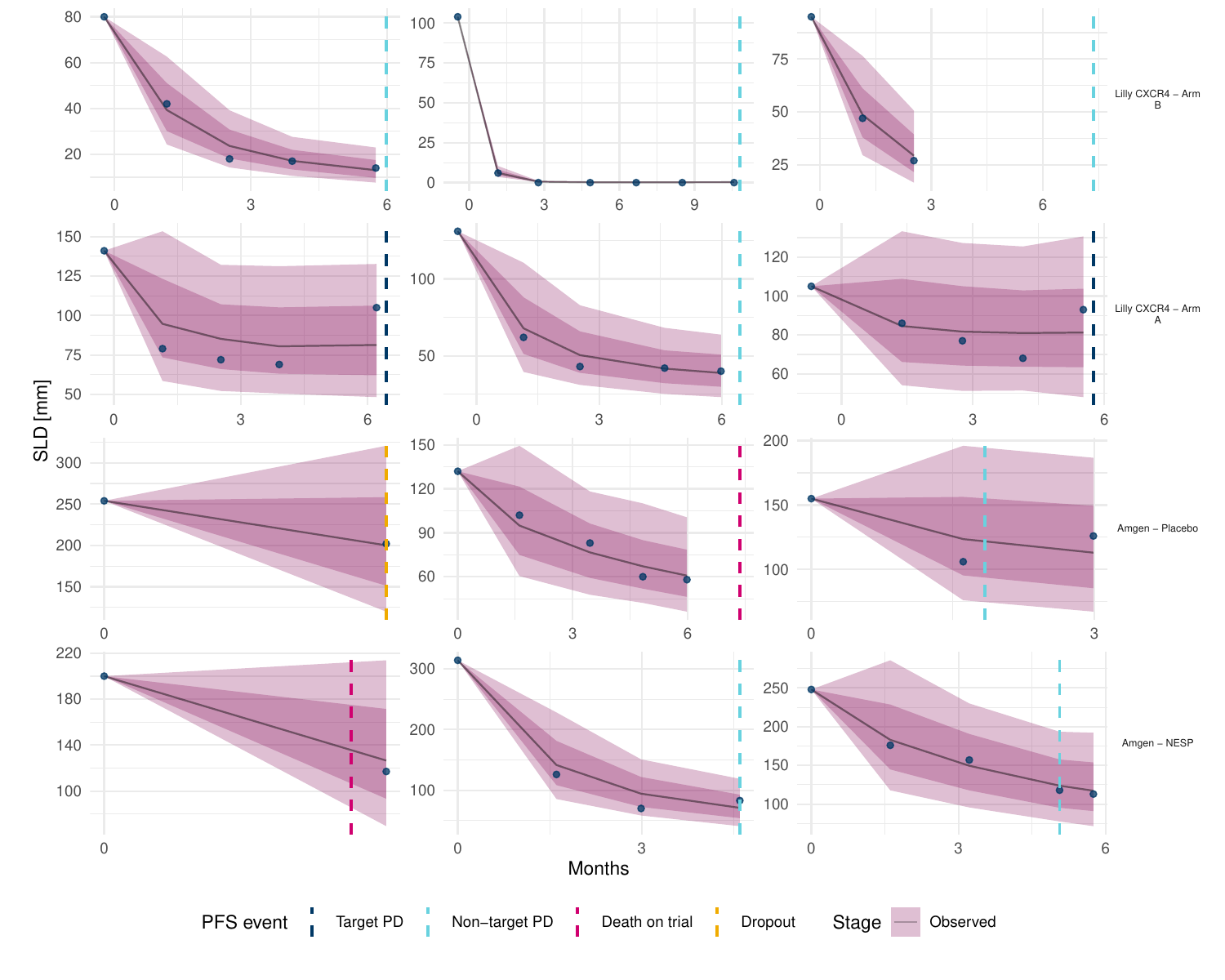}}

}

\caption{\label{fig-sld-ppc}\textbf{SCLC: tumour-burden fit for
progressed patients.} Posterior-predictive SLD trajectories for a random
subsample of patients with observed progression, drawn across the Lilly
CXCR4 target arms and the Amgen Darbe historical arms. Ribbons represent
50\% and 80\% credible intervals; lines show posterior median tumour
dynamics. Dots are observed SLD measurements. Vertical dashed lines
indicate the week of the PFS event.}

\end{figure}

Figure~\ref{fig-sld-ppc} shows posterior predictive SLD trajectories for
a random subsample of patients with observed progression events. For
each patient, the posterior median trajectory (solid line) tracks the
observed SLD measurements (dots), with 50\% and 80\% credible ribbons
capturing the measurement uncertainty; vertical dashed lines mark the
PFS event week. The two-component decomposition correctly recovers the
characteristic shapes: initial response (driven by the decreasing
component \(D_i\)) followed by eventual regrowth (driven by the growing
component \(G_i\)). The width of credible intervals varies across
patients even when observation counts are similar, reflecting three
sources of patient-level uncertainty: the number and temporal span of
observations available to identify the per-patient kinetic parameters;
the geometric identifiability of the trajectory shape (a clear nadir
constrains parameters better than a monotone decline); and the degree of
agreement between a patient's kinetics and the population distribution
--- patients whose trajectories are close to the hierarchical mean
receive stronger regularisation from the prior and exhibit tighter
posteriors, while patients with atypical kinetics show wider intervals
as the model negotiates between their sparse data and the population.
Overall, the posterior median trajectories track the observed
measurements closely across all patients, confirming that the
two-component model provides an adequate fit to the longitudinal tumour
data.

\begin{figure}

\centering{

\pandocbounded{\includegraphics[keepaspectratio]{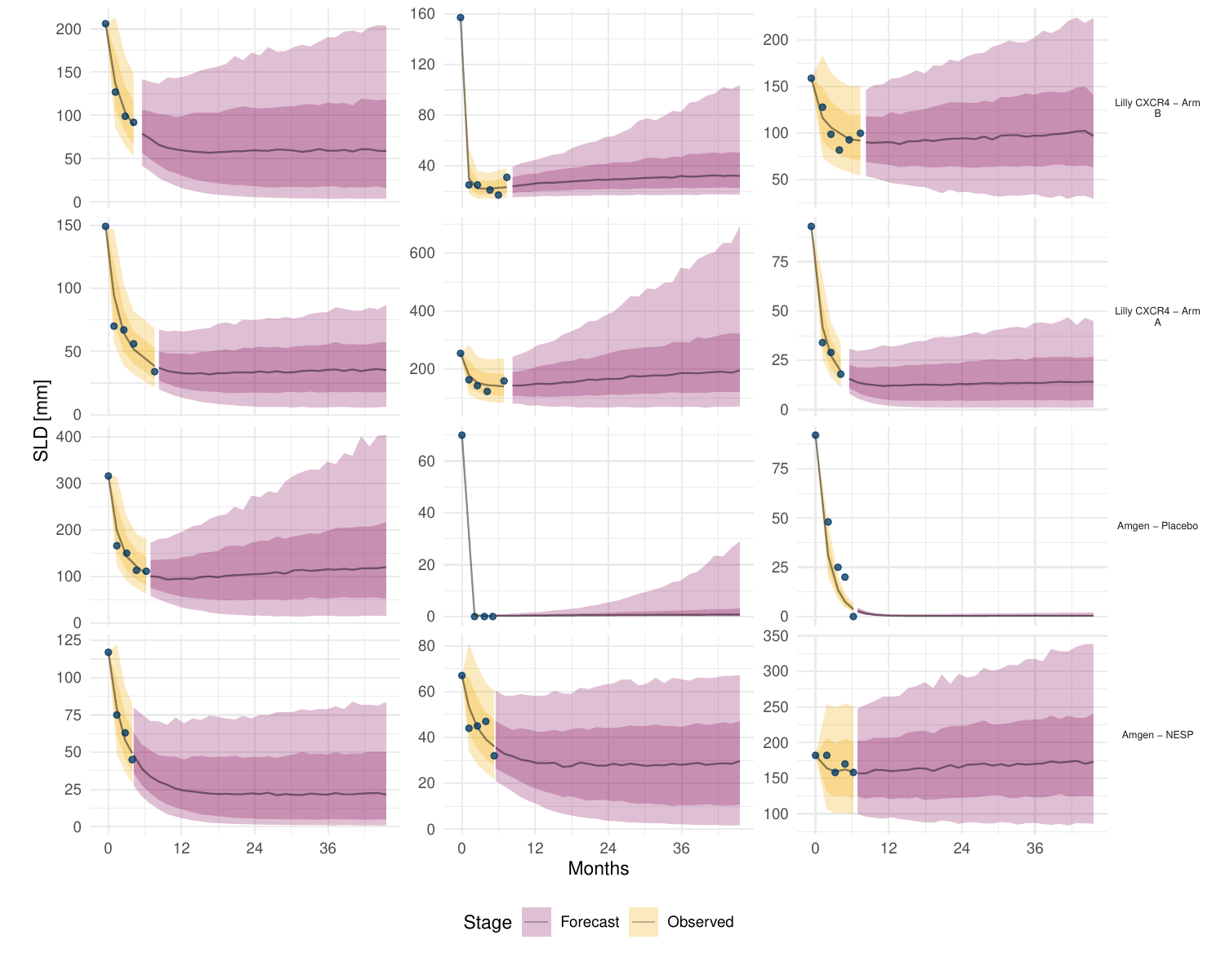}}

}

\caption{\label{fig-sld-censored}\textbf{SCLC: tumour-burden forecast
for censored patients.} Posterior-predictive SLD trajectories for a
subsample of censored patients (progression-free at data cut-off), drawn
across the Lilly CXCR4 target arms and the Amgen Darbe historical arms.
Dark ribbons show the observed phase; pink ribbons show the forecast
beyond the last observation.}

\end{figure}

Figure~\ref{fig-sld-censored} shows the same posterior predictive check
for censored patients (those still progression-free at the data
cut-off). Here the model must \emph{forecast} beyond the last
observation. The forecast ribbons (pink) fan out naturally from the
observed phase (dark), reflecting increasing uncertainty further from
the data while maintaining the structural decrease-plus-growth shape.
This is the extrapolation that makes mechanistic modelling valuable: the
model predicts \emph{what will happen} based on the inferred rates, not
just \emph{what has happened}.

\begin{figure}

\centering{

\pandocbounded{\includegraphics[keepaspectratio]{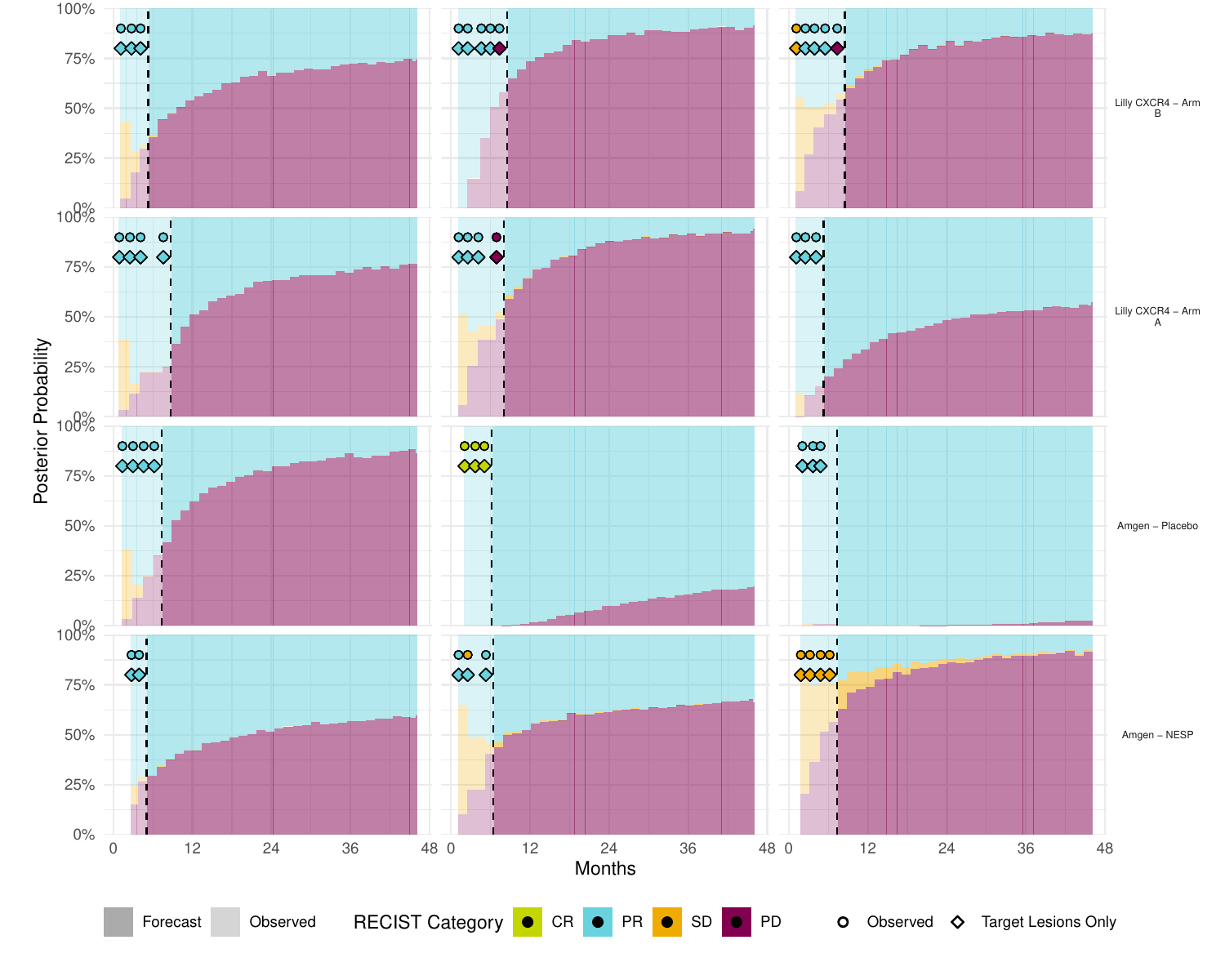}}

}

\caption{\label{fig-recist-censored}\textbf{SCLC: RECIST-category
forecast for censored patients.} Posterior-predictive RECIST
classification over time for a subsample of censored patients, drawn
across the Lilly CXCR4 target arms and the Amgen Darbe historical arms.
Stacked bars show the posterior probability of each RECIST category at
each time point. Coloured dots represent observed classifications;
dashed lines mark the transition from observed to forecast phase.}

\end{figure}

Figure~\ref{fig-recist-censored} shows the posterior predictive RECIST
classification over time for the same censored patients. Each stacked
bar represents the posterior probability of each RECIST category (CR,
PR, SD, PD) at each time point. The observed RECIST classifications
(coloured dots) are consistent with the high-probability regions, and
the gradual shift from PR/SD toward PD in the forecast window reflects
the growing component's eventual dominance.

\subsection{Forecasting mature survival from immature
data}\label{sec-results-lfo}

The central claim of PIONEER is that it can forecast survival endpoints
from immature data. We evaluate this claim using leave-future-out (LFO)
cross-validation: the model is re-fitted at a sequence of progressively
later data cut-offs (month 4, 11, and 19 of enrolment), and at each
cut-off the predicted PFS and OS Kaplan--Meier curves are compared
against the observed mature-DCO KM that was \emph{not} available at
training time.

The forecasts shown here are the model's \emph{conditional} predictions
(the \(\mathrm{sample}\) estimand of Section~\ref{sec-ms-forecast}): for
each patient the model retains the event history actually observed up to
the cut-off --- who had already progressed, died, or dropped out, and
who was still progression-free --- and re-simulates only the unobserved
remainder of their trajectory. This is deliberate. The conditional
forecast answers the operational question a trialist faces at an interim
look --- \emph{``given everything we know about these patients today,
what will the mature KM look like?''} --- and is the like-for-like
counterpart to the cut-off-censored observed KM it is compared against.
It is distinct from the \emph{unconditional} (\(\mathrm{spop}\))
forecast used for the population calibration check in
Section~\ref{sec-results-survival}, which discards all observed
histories and re-simulates every patient from treatment start; that
estimand answers a different question (can the model regenerate the
entire KM from population parameters alone?) and is the wrong object for
a forecasting evaluation, because in practice the patient-level data
accumulated up to the cut-off \emph{is} known and ignoring it would
discard the very information that makes early forecasting possible.

\begin{landscape}

\begin{figure}

\centering{

\pandocbounded{\includegraphics[keepaspectratio]{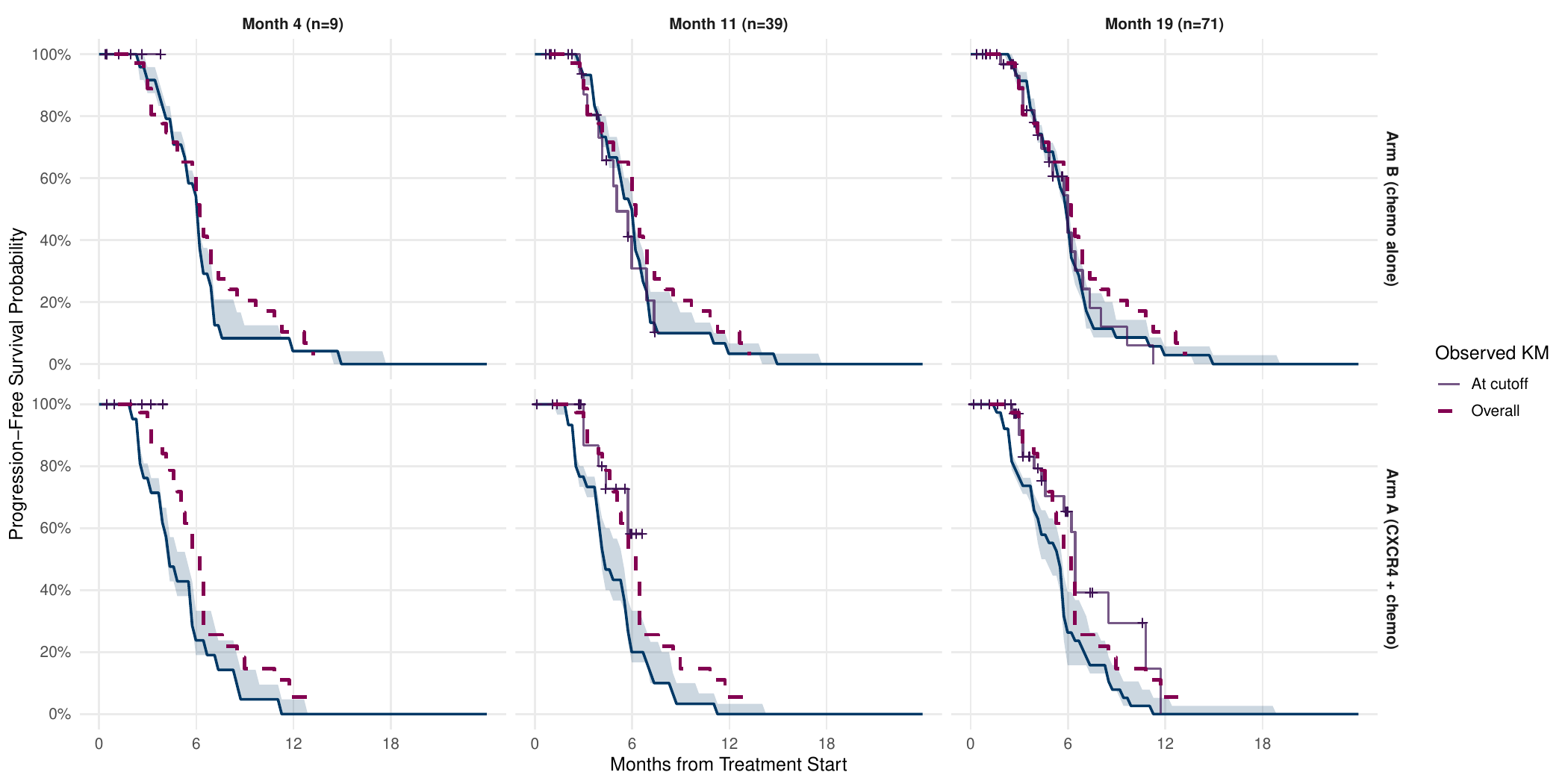}}

}

\caption{\label{fig-lfo-pfs-evolution}\textbf{SCLC: leave-future-out PFS
forecast evolution.} Conditional forecast for the Lilly CXCR4 trial:
each patient's observed event history up to the cut-off is retained and
only the unobserved remainder is re-simulated. Each column is a data
cut-off; rows are treatment arms. Navy ribbons show 80\% posterior
credible intervals for the conditional forecast; navy line shows the
posterior median. The solid step line shows the observed PFS KM censored
at the cut-off date; the dashed line shows the final mature observed PFS
KM as a reference (identical across columns). The forecast narrows and
converges toward the mature curve as data accumulates.}

\end{figure}%

\end{landscape}

Figure~\ref{fig-lfo-pfs-evolution} shows the PFS forecast evolution for
the two arms of the Lilly CXCR4 trial. At the earliest cut-off (month 4,
\(n=9\)), the model has seen only a handful of patients and produces
wide credible intervals --- but the median prediction already captures
the rapid early decline characteristic of this population. As more
patients and longer follow-up become available (month 11, \(n=39\);
month 19, \(n=71\)), the credible intervals narrow and the predicted KM
converges toward the observed mature-DCO curve. For Arm B, the forecast
is well-calibrated at all cut-offs. For Arm A, however, the posterior
predictive median remains systematically pessimistic --- it tracks
closer to the mature ``Overall'' KM than to the ``At cutoff'' observed
data on which the model was trained. The model predicts earlier
progression than the cutoff-censored KM suggests, even though the
prediction interval is narrow. This overconfident pessimism reflects the
growth-rate identifiability issue discussed in
Section~\ref{sec-discussion}: the mechanistic component predicts
target-lesion RECIST progression based on a population growth rate that
is biased by a subset of fast-progressing patients, and patients with
insufficient follow-up to identify their own growth rate are shrunk
toward this pessimistic estimate. Despite this, the forecasts remain
informative --- the predicted median PFS is close to the final mature
value, suggesting that the model correctly anticipates the
\emph{eventual} outcome even when it is overconfident about
\emph{timing} at early cut-offs.

\begin{landscape}

\begin{figure}

\centering{

\pandocbounded{\includegraphics[keepaspectratio]{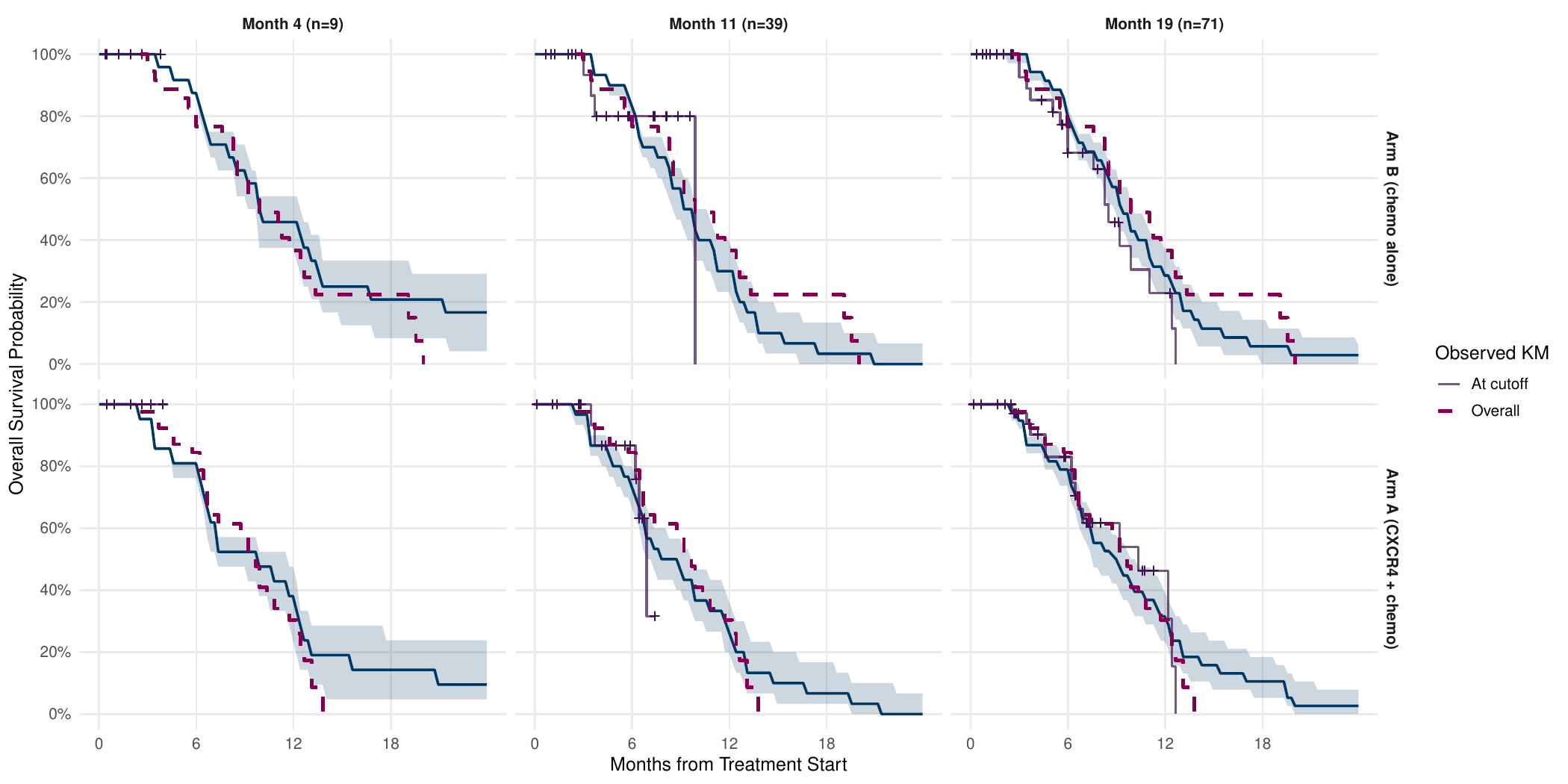}}

}

\caption{\label{fig-lfo-os-evolution}\textbf{SCLC: leave-future-out OS
forecast evolution.} Conditional forecast for the Lilly CXCR4 trial;
same structure as Figure~\ref{fig-lfo-pfs-evolution} but for overall
survival. The model conditions on observed deaths and censoring up to
each cut-off and forecasts OS for patients still alive. Wider credible
intervals reflect the additional uncertainty from the full multistate
routing through progression, direct death, and dropout.}

\end{figure}%

\end{landscape}

Figure~\ref{fig-lfo-os-evolution} shows the corresponding OS forecast
evolution. OS prediction is inherently harder --- it requires the full
multistate routing through progression, dropout, and post-progression
death --- yet the model's predictions remain calibrated across cut-offs.
The wider ribbons relative to PFS reflect the additional structural
uncertainty from the competing transitions, not miscalibration.

\clearpage

\subsection{Clinical endpoint summaries}\label{sec-results-endpoints}

Having shown the full forecast KM curves converging across cut-offs
(Section~\ref{sec-results-lfo}), we now read off the scalar clinical
endpoints that those forecasts imply --- median PFS, median OS, and ORR
--- the headline numbers a trial actually reports. These summaries use
the same conditional (\(\mathrm{sample}\)) estimand and the same
sequence of LFO cut-offs, so they tell the forecasting story in the
quantities decision-makers quote rather than as full survival curves.
The full-posterior propagation delivers each as a distribution rather
than a point estimate, making explicit the uncertainty that classical
point estimators suppress. For each endpoint, the prior distribution
spans the range of outcomes consistent with the biological constraints
and the historical population before the target trial is observed; the
posterior concentrates that distribution using the target-trial data.

\begin{figure}

\centering{

\pandocbounded{\includegraphics[keepaspectratio]{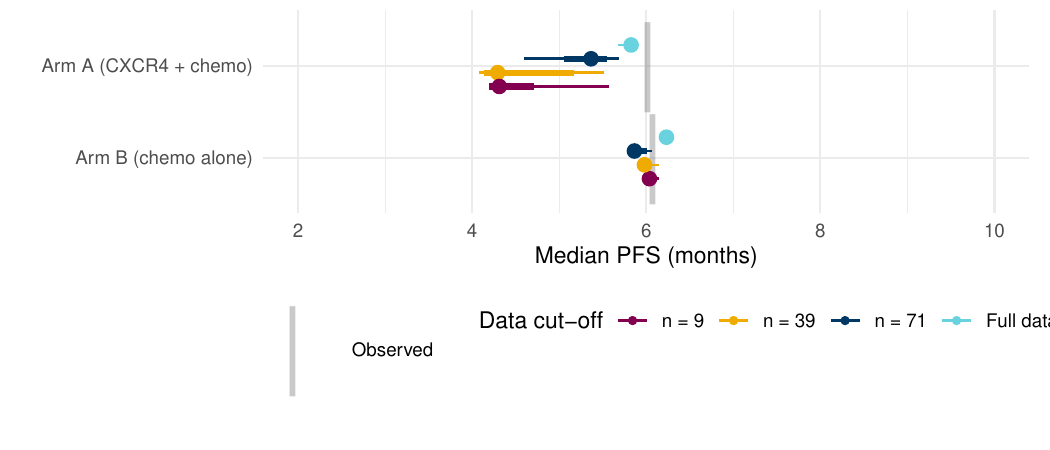}}

}

\caption{\label{fig-median-pfs}\textbf{SCLC: median PFS forecast across
cut-offs.} Posterior median PFS (months) by arm (Lilly CXCR4) at each
LFO cut-off and the final full-data fit. Colours indicate progressive
data cut-offs by number of target-trial patients observed. Points show
posterior medians; thick bars = 50\% CI, thin lines = 90\% CI. Vertical
bar = observed mature-DCO value.}

\end{figure}

Figure~\ref{fig-median-pfs} shows the posterior distribution of median
PFS evolving across LFO cut-offs for the Lilly CXCR4 trial arms. As more
patients are observed, the posterior median narrows and converges toward
the observed mature value (dashed line, \textasciitilde6.2 months).
Median PFS is a joint-model endpoint --- it combines both the
mechanistic target-lesion progression and the stochastic multistate
progression channels.

\begin{figure}

\centering{

\pandocbounded{\includegraphics[keepaspectratio]{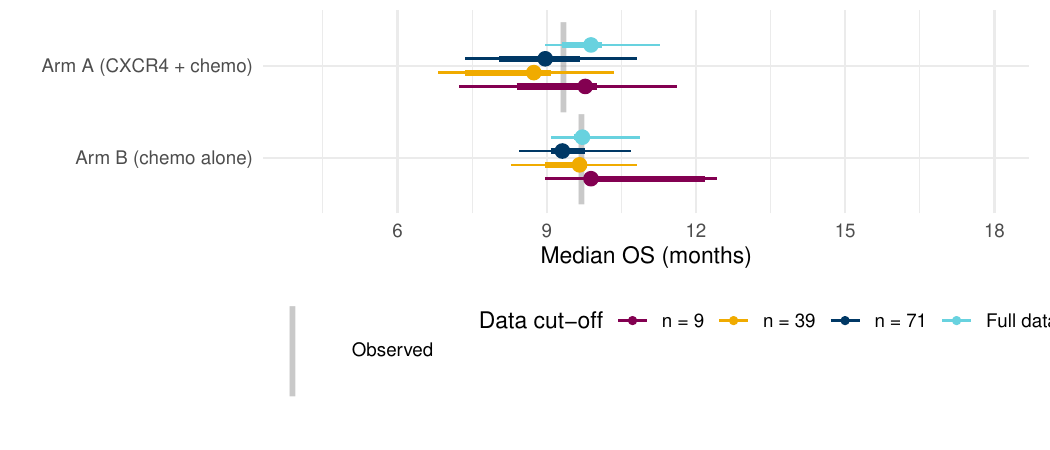}}

}

\caption{\label{fig-median-os}\textbf{SCLC: median OS forecast across
cut-offs.} Posterior median OS (months) by arm (Lilly CXCR4) at each LFO
cut-off and the final full-data fit. Colours indicate progressive data
cut-offs by number of target-trial patients observed. Points show
posterior medians; thick bars = 50\% CI, thin lines = 90\% CI. Vertical
bar = observed mature-DCO value.}

\end{figure}

Figure~\ref{fig-median-os} shows the corresponding OS median evolution.
OS convergence is slower than PFS, reflecting the additional structural
uncertainty from multistate routing, but the model's predictions remain
centred on the observed mature value across cut-offs.

\begin{figure}

\centering{

\pandocbounded{\includegraphics[keepaspectratio]{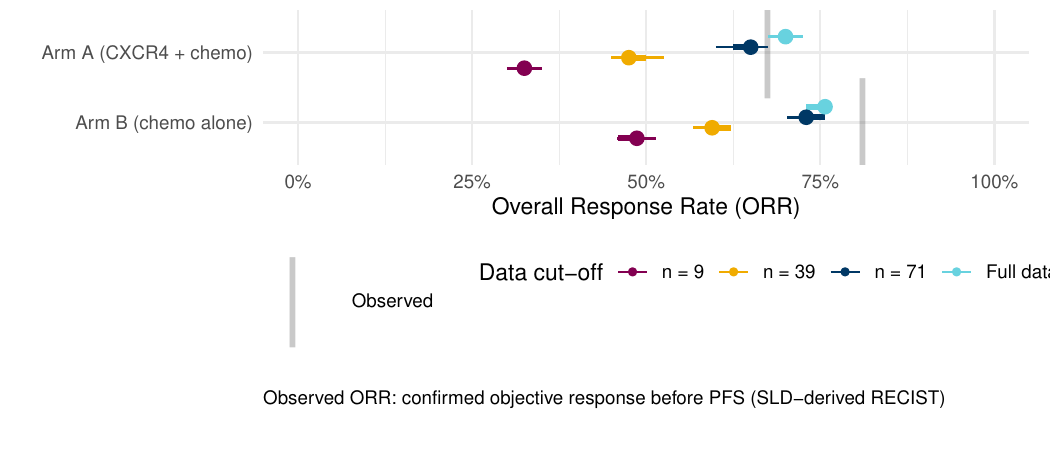}}

}

\caption{\label{fig-orr}\textbf{SCLC: ORR forecast across cut-offs.}
Posterior ORR by arm (Lilly CXCR4) at each LFO cut-off and at the final
full-data fit. Colours indicate progressive data cut-offs by number of
target-trial patients observed. ORR is derived entirely from the
mechanistic submodel. Points show posterior medians; thick bars = 50\%
CI, thin lines = 90\% CI. Vertical bar = observed ORR from the clinical
study report.}

\end{figure}

Figure~\ref{fig-orr} shows the ORR posterior by arm across LFO cut-offs.
Unlike median PFS, ORR is derived \textbf{entirely from the mechanistic
submodel} --- it is a deterministic function of the latent SLD
trajectory (whether \(B_i(t)\) ever falls below 0.7, i.e.~a
\(\geq 30\%\) decrease from baseline). No multistate hazard is involved.
This is why the ORR posterior is remarkably tight even at early cut-offs
--- response depth is well-identified from even a few SLD observations,
and the structural decrease-plus-growth model constrains the trajectory
shape. A hazard-only survival model without the mechanistic component
could not produce ORR predictions at all.

\subsection{Population calibration: unconditional posterior-predictive
checks}\label{sec-results-survival}

The forecasts in Section~\ref{sec-results-lfo} are the model's intended
use; this section is a complementary goodness-of-fit diagnostic. We
assess the model's population-level calibration using an
\emph{unconditional} posterior predictive check: every patient's event
time is re-simulated from week 1, discarding all observed event
histories. The model must reconstruct the full Kaplan--Meier curve from
the inferred population parameters, patient-level covariates, and latent
tumour trajectories alone. Unlike the conditional forecast --- which is
anchored to each patient's observed history and is the object we
actually predict with --- this unconditional pass is not a forecast but
a check: if it recovers the mature observed KM from the population
machinery alone, the model has correctly learned the generative
distribution rather than memorised outcomes.

This is a calibration check, not a forecasting evaluation. The
operationally relevant forecasting question --- ``given the data
available today, what will the final KM look like?'' --- was addressed
in Section~\ref{sec-results-lfo} above, where the model is re-fitted at
progressively earlier cut-offs and its predictions are compared against
the mature KM that was not yet available at training time.

\begin{figure}

\centering{

\pandocbounded{\includegraphics[keepaspectratio]{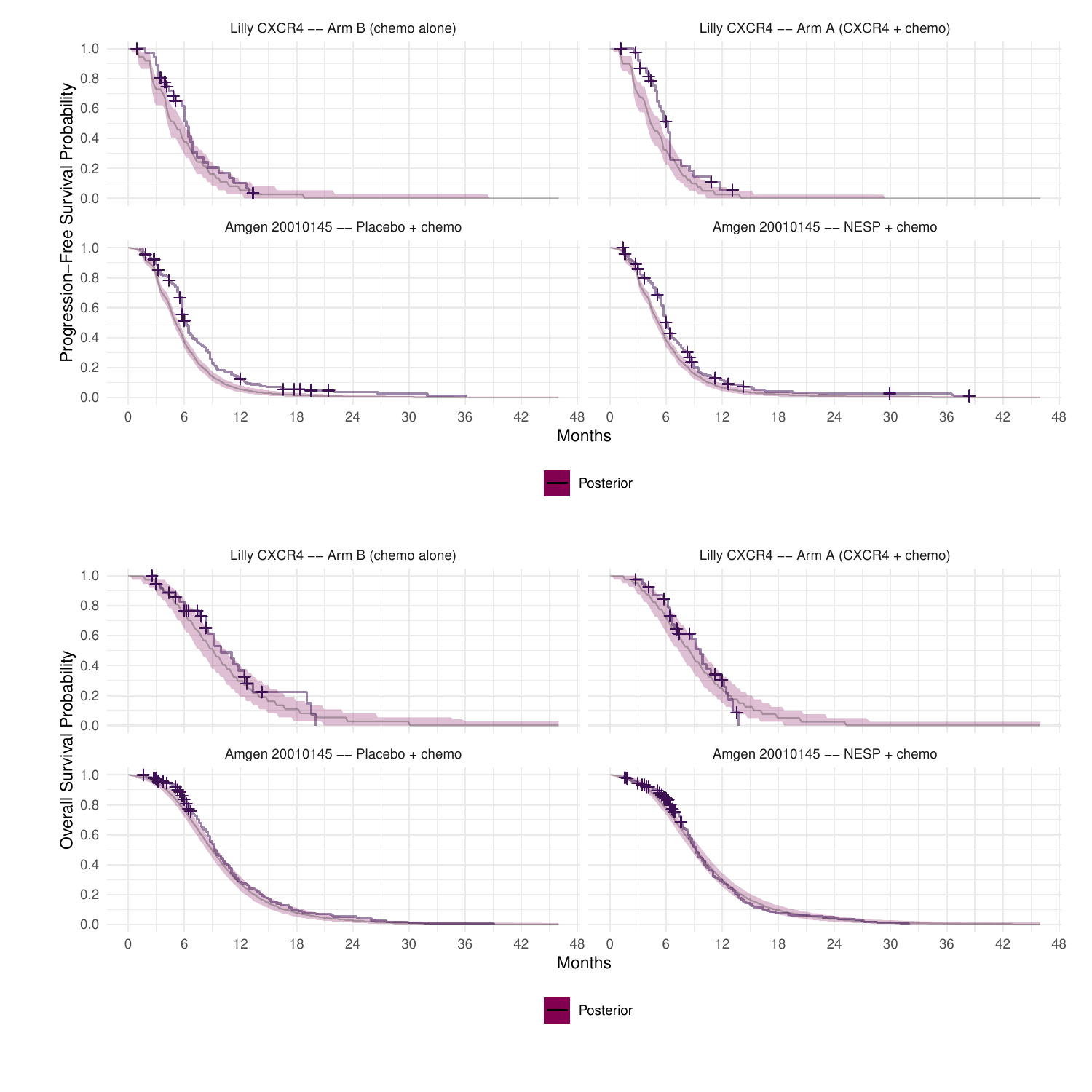}}

}

\caption{\label{fig-km-spop}\textbf{SCLC: unconditional KM calibration
check.} Every patient is re-simulated from week 1; no observed events
are used. Ribbons show 80\% credible intervals; step lines show the
observed KM. \textbf{Top:} PFS by trial and arm. \textbf{Bottom:} OS by
trial and arm.}

\end{figure}

Figure~\ref{fig-km-spop} shows the unconditional posterior predictive
check for PFS (top) and OS (bottom) by trial and arm. The posterior
median PFS curve tracks the observed KM closely across all four arms,
and the 80\% credible ribbon covers the observed step function
throughout. The OS check requires the full multistate routing --- a
patient's OS depends on which path they take through the state graph
(direct death via \(0\to 2\), post-progression death via \(1\to 2\), or
dropout via \(3\to 2\)) --- yet the model recovers the observed OS KM
well, with credible intervals that are wider than PFS, as expected from
the additional uncertainty propagated through the competing transitions.

\subsection{Covariate effects and tumour-hazard
coupling}\label{sec-results-covariates}

\begin{figure}

\centering{

\pandocbounded{\includegraphics[keepaspectratio]{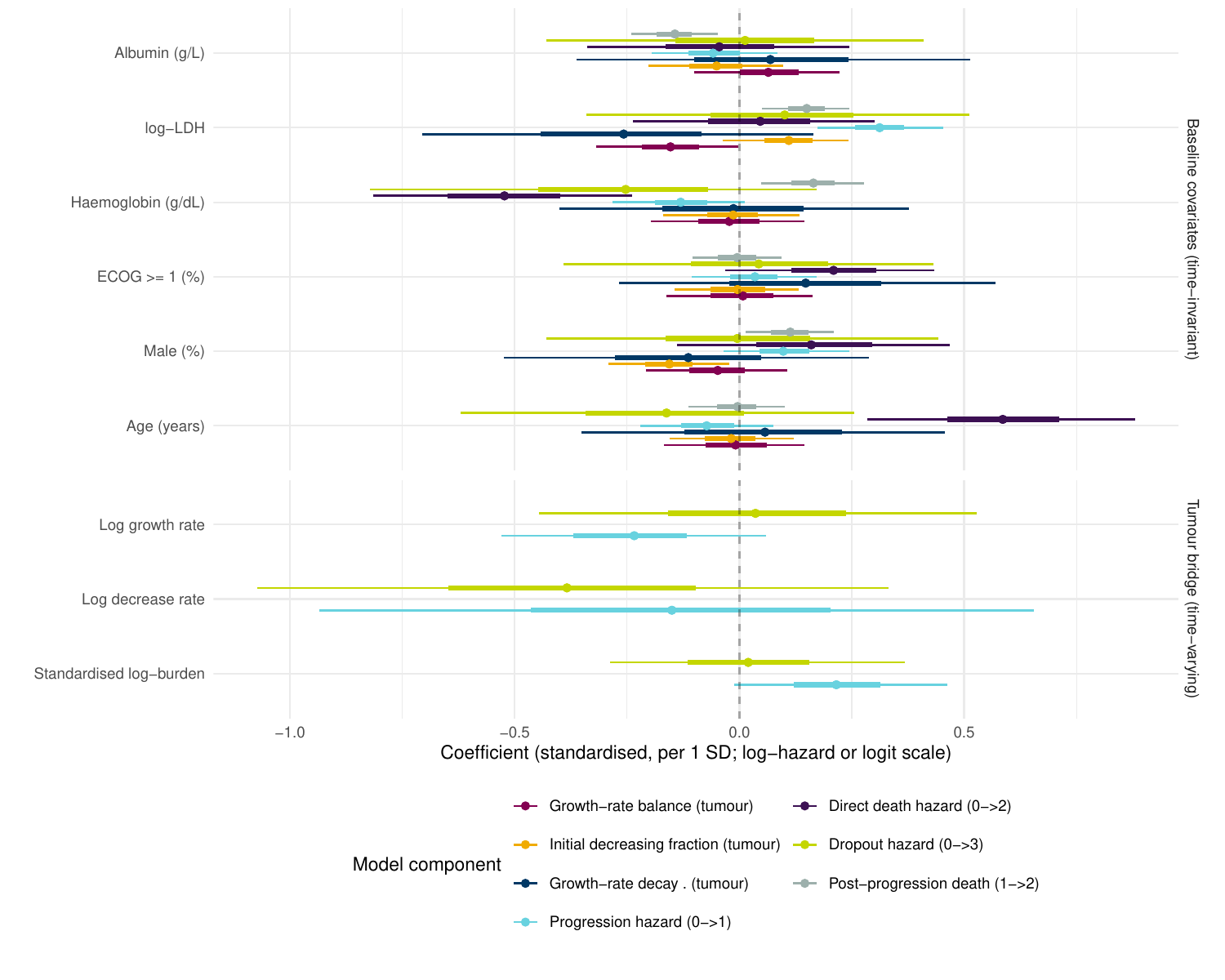}}

}

\caption{\label{fig-covariate-effects}\textbf{SCLC: posterior covariate
effects across model components.} \textbf{Top panel --- time-invariant
baseline covariates} (back-transformed from QR reparameterisation):
growth-rate balance, initial decreasing fraction, Gompertz decay
\(\kappa\) (tumour dynamics), and four multistate transition hazards.
\textbf{Bottom panel --- time-varying tumour bridge} \(\mathbf{W}_i(t)\)
(Section~\ref{sec-ms-bridge}): standardised latent log-burden and log
decrease/growth rates for the two bridge-wired transitions (\(0\to 1\),
\(0\to 3\)). Both panels are on a per-1-SD scale (log-hazard or logit).
Intervals show 50\% and 90\% credible intervals.}

\end{figure}

Figure~\ref{fig-covariate-effects} shows the posterior distribution of
covariate coefficients across all model components: the tumour-dynamics
submodel (growth-rate balance, initial decreasing fraction, Gompertz
decay \(\kappa\)), the four multistate transition hazards (\(0\to 1\)
progression, \(0\to 2\) direct death, \(0\to 3\) dropout, \(1\to 2\)
post-progression death), and the time-varying tumour bridge
(Section~\ref{sec-ms-bridge}). All coefficients are on a standardised,
per-one-standard-deviation scale, back-transformed from the QR
reparameterisation used during sampling.

The key observation is that the same covariate can act in opposite
directions on different components. Haemoglobin, for example, has a
clear negative coefficient on the \(0\to 1\) progression hazard (higher
Hgb → lower progression risk) but a positive coefficient on \(1\to 2\)
post-progression death --- consistent with anaemia being prognostic for
both events through different mechanisms. The \(0\to 2\) direct-death
coefficients have the widest credible intervals, reflecting the small
number of direct-death events. This transition-specific resolution is a
structural advantage of the multistate formulation over a single-hazard
model, which would collapse these opposing effects into one coefficient.

The bottom panel shows the time-varying bridge \(\mathbf{W}_i(t)\) ---
standardised latent log-burden and the patient-specific log
decrease/growth rates. The bridge feeds two transitions in the SCLC
configuration (\(0\to 1\) and \(0\to 3\)); direct death and
post-progression death carry no bridge coefficient here. The positive
log-burden coefficient on \(0\to 1\) is the mechanistic heart of the
joint model: the stochastic progression hazard rises as the latent
tumour burden grows. Decrease/growth-rate coefficients are more diffuse
--- burden level carries most of the per-visit signal once conditioned
on --- and the dropout bridge is identified only weakly.

\section{Discussion}\label{sec-discussion}

The primary contribution of PIONEER is not a better fit to historical
data --- a sufficiently flexible semiparametric model imposes fewer
constraints on the trajectory space and therefore cannot achieve worse
in-sample fit than PIONEER when given comparable or greater effective
degrees of freedom --- but rather the ability to produce calibrated
population-level forecasts from immature data, at a point in a trial's
life cycle when the clinical decision (continue, expand, terminate) must
be made but the events have not yet occurred. The structural advantage
lies not in fitting observed data more closely but in constraining the
extrapolation. The value proposition is therefore one of \emph{decision
acceleration}: bringing the information content of a mature data cut-off
forward in time by imposing biologically informed structural constraints
on the extrapolation.

In drug development, the cost of a late-stage failure dwarfs the cost of
early-stage programmes. A Phase III trial that reads out futility at an
interim analysis has already consumed years of enrolment and hundreds of
millions of dollars; a Phase II programme that could have been killed
--- or accelerated --- six months earlier on the basis of tumour-kinetic
evidence would have freed those resources for more promising compounds.
PIONEER operationalises this logic: the leave-future-out validation
(Section~\ref{sec-results-lfo}) demonstrates that the joint model's
forecasts are informative at early data cut-offs where the majority of
target PFS events have not yet occurred. When embedded within a go/no-go
decision framework, this translates into earlier and better-informed
decisions about programme continuation --- with the conservative bias of
the mechanistic channel providing a natural safety margin against
advancing compounds that will ultimately fail.

\subsection{Growth-rate identifiability and conservative PFS
forecasting}\label{growth-rate-identifiability-and-conservative-pfs-forecasting}

The leave-future-out analysis reveals a structural asymmetry in the
mechanistic submodel's forecasting performance. The growth rate
\(r^{\mathrm{gro}}_i\) is the least identifiable patient-level
parameter, because the growing compartment contributes negligibly to the
observed SLD until regrowth is underway. For patients whose follow-up
ends before visible regrowth --- the majority at early data cut-offs ---
the patient-level growth-rate posterior remains wide and is dominated by
the hierarchical population estimate. The population growth rate is, in
turn, disproportionately informed by patients who \emph{have} reached
their nadir and begun regrowing: these are the patients for whom the
growing compartment is non-negligible in the likelihood and whose data
actually identifies \(r^{\mathrm{gro}}_i\). Among these identifiable
patients, faster growers contribute more information per unit follow-up
because the growing compartment's signal-to-noise ratio scales with its
magnitude relative to the decreasing compartment. The population
estimate is therefore pulled toward the upper tail of the true
growth-rate distribution --- though we note that alternative
explanations (e.g.~patients reaching nadir early due to fast decrease
rather than fast growth) cannot be excluded and would attenuate rather
than reverse this bias.

Hierarchical shrinkage then acts asymmetrically: patients with poorly
identified growth rates --- those still in their response phase --- are
pulled toward this population estimate, producing predictions that are
systematically more pessimistic than the eventual observed PFS. The
model is not merely uncertain; it is \emph{overconfident} in a
pessimistic direction, because the population rate is precisely
estimated even though it may not represent the full patient population.
The leave-future-out validation quantifies both the direction and
magnitude of this conservative bias, establishing that it is systematic
rather than random: the predictive interval sits consistently below the
observed PFS Kaplan--Meier (Figure~\ref{fig-lfo-pfs-evolution}). The
sensitivity of this bias to the population-level prior on
\(r^{\mathrm{gro}}\) merits investigation in future work; in the present
analysis, the weakly informative prior is dominated by the data for arms
with sufficient post-nadir follow-up.

We note that this conservative bias is not necessarily a deficiency in
the intended use case. In a go/no-go decision framework, a model that is
pessimistic about PFS provides a natural safety margin: if the forecast
meets the decision threshold despite its conservative tendency, the true
outcome is likely to exceed it. The alternative --- an optimistically
biased model --- carries the more dangerous risk of advancing a
programme that will ultimately fail. The asymmetric cost of false
continuation (advancing a failing programme, consuming years of
downstream investment) versus false termination (stopping a viable
programme, expensive but recoverable) means that conservative bias is
aligned with the decision-theoretic incentives of the framework. This
framing does not excuse miscalibration --- a perfectly calibrated model
would be strictly preferable --- but it contextualises the direction of
the observed bias as less costly than the alternative.

\subsection{Differential calibration of PFS and
OS}\label{differential-calibration-of-pfs-and-os}

The unconditional posterior-predictive checks
(Section~\ref{sec-results-survival}; Figure~\ref{fig-km-spop}) show that
OS forecasts are better calibrated than PFS forecasts across both
trials. This difference has a mechanistic explanation rooted in the
model architecture, though differential event counts and interval
censoring may also contribute. PFS depends heavily on the mechanistic
channel: progression is triggered when the latent SLD trajectory crosses
the RECIST threshold, which is directly determined by the patient's
growth rate --- the very parameter that suffers from the identifiability
asymmetry described above. OS, by contrast, is primarily driven by the
multistate hazard channel: overall survival integrates over the
stochastic transition hazards (progression-to-death, direct death),
which are estimated from observed event data rather than extrapolated
from the mechanistic trajectory. The hazard parameters do not suffer
from the same directional identifiability bias because they are informed
by the occurrence and timing of events --- data that is present
regardless of whether the patient has reached their SLD nadir.

The multistate-only PFS --- which triggers progression from the
stochastic hazard rather than from the RECIST threshold crossing ---
confirms this explanation: it does not exhibit the same systematic
pessimism, because it bypasses the growth-rate extrapolation entirely.
The implication is that for arms or subgroups where most patients have
short, monotonically declining SLD trajectories at the data cut-off, the
hazard channel provides more reliable PFS forecasts than the mechanistic
channel alone.

\subsection{Individual-level predictions and clinical decision
support}\label{individual-level-predictions-and-clinical-decision-support}

Beyond trial-level forecasting, the joint posterior furnishes
patient-specific predictive distributions. For any patient \(i\) with
data observed through week \(t\), the model produces a full posterior
predictive distribution
\(P(\text{progression by week } T \mid \text{data}_i(1:t))\) --- not a
point estimate but a probability statement that updates as new
assessments arrive. This individual-level output has natural
applications in clinical decision support: an oncologist reviewing a
patient's trajectory could, in principle, query the model for the
probability of progression within the next assessment cycle, conditional
on everything observed so far.

We emphasise that PIONEER has not been prospectively validated for
individual clinical decision-making, and the results presented here
constitute a methodological demonstration rather than a clinical tool
evaluation. Population-level calibration --- which we demonstrate here
--- is a necessary but not sufficient condition for individual-level
probability statements to be well-calibrated; the latter requires
assessment of calibration at the patient level in a held-out cohort, and
integration into clinical workflows where the probabilistic output can
inform --- but not replace --- clinician judgment. Nevertheless, the
architecture is designed to support this use case, and the path from
methodological framework to clinical tool is one of validation rather
than re-engineering.

\subsection{Limitations}\label{limitations}

Several limitations merit discussion. First, the decrease-plus-growth
decomposition is phenomenological: the two components are not claimed to
recover distinct clonal populations, and the model cannot distinguish
true clonal heterogeneity from, for example, a change in drug
sensitivity over time. The case for the decomposition rests on
forecasting performance and parameter transportability, not on a
biological mechanism claim. Second, the multistate submodel's
identifiability depends on having sufficient events per transition; in
small trials or in arms with very few direct deaths, some
transition-specific parameters will be weakly identified and dominated
by the prior. Third, computational cost remains substantial: a single
model fit requires several hours on modern hardware, limiting the
feasibility of rapid iterative analyses --- though this cost is paid
once per data cut-off rather than per endpoint. Fourth, the model does
not improve in-sample fit relative to flexible empirical alternatives; a
sufficiently flexible semiparametric model (e.g.~Gaussian process or
spline-based) imposes fewer constraints on the trajectory space and
therefore cannot achieve worse in-sample fit given comparable effective
degrees of freedom. The advantage of PIONEER is entirely in
out-of-sample extrapolation under structural constraints. Fifth, the
hierarchical shrinkage behaviour described above means that the model's
forecasting properties depend on the composition of the population-level
prior and the amount of post-nadir data available --- a dependence that,
while standard for hierarchical Bayesian models, should be acknowledged
as a sensitivity.

\subsection{When to prefer simpler
models}\label{when-to-prefer-simpler-models}

PIONEER's complexity is justified only when its structural assumptions
earn their keep in forecasting accuracy. A simpler model --- parametric
survival extrapolation, or even non-parametric Kaplan--Meier with
confidence bands --- may be preferable when: (i) follow-up is already
mature and the extrapolation horizon is short; (ii) the target endpoint
is OS only and the mechanistic dynamics add no information beyond what
the event times already provide; (iii) the trial is small enough that
the multistate submodel's parameters cannot be identified separately
from the population prior; or (iv) the decision does not require
individual-level predictions and population-level summaries suffice. The
go/no-go decision threshold itself is a derived quantity that PIONEER
enables, but the model's full capability --- individual trajectories,
transition-specific hazards, cross-trial borrowing --- is not required
for every decision context.

\subsection{Generalisability}\label{generalisability}

The framework has been applied to other continuous longitudinal
biomarkers with analogous decrease-plus-growth dynamics. This
demonstrates that the architecture is not specific to SLD or
RECIST-defined endpoints; the mechanistic submodel generalises to any
setting where the observed biomarker can be decomposed into a responding
and a refractory component, and where the multistate transitions of
clinical interest can be linked to the latent trajectory through
time-varying covariates. The population hierarchy and the bridge
specification are modular: they accommodate different biomarker scales,
different transition structures, and different covariate vectors without
architectural modification.

A natural question is whether cross-indication borrowing --- fitting the
model jointly across trials from different tumour types --- can improve
forecasts for a new indication with limited data. The current case study
pools trials within indication; whether the population-level kinetic
parameters (growth rates, response fractions, hazard baselines) are
sufficiently portable across indications to justify cross-indication
hierarchies remains an open empirical question that we do not address
here.

\subsection{Future directions}\label{future-directions}

Several extensions are natural. First, incorporating additional
longitudinal biomarkers (e.g.~ctDNA, LDH) as co-modelled latent
processes --- rather than as fixed baseline covariates --- would allow
the model to exploit real-time information that may precede radiological
response. Second, integration with model-based meta-analysis would
formalise the cross-trial borrowing and enable Bayesian design of Phase
III trials informed by the posterior predictive distribution from Phase
II --- leveraging the structural model's ability to forecast endpoints
from early kinetic data to sharpen design assumptions for late-stage
trials. Third, real-time deployment for adaptive trial monitoring, where
the model is re-fitted at each data transfer and the go/no-go threshold
is evaluated prospectively, is technically feasible given the current
computational pipeline. Fourth, regulatory positioning within the FDA
Model-Informed Drug Development (MIDD) framework is a natural direction;
the combination of mechanistic interpretability, uncertainty
quantification, and demonstrated out-of-sample calibration aligns with
the properties that regulators have identified as supportive of
supplementary evidence from model-based analyses (U.S. Food and Drug
Administration 2021).

\appendix

\section*{Appendix A. Target-lesion vs clinical
endpoints}\label{sec-appendix-target}
\addcontentsline{toc}{section}{Appendix A. Target-lesion vs clinical
endpoints}

The mechanistic submodel (Section~\ref{sec-mechanistic}) sees only the
sum of target-lesion diameters. Clinical RECIST and clinical PFS, by
contrast, also respond to non-target lesions, new lesions, and death ---
events that leave no signature in the target-lesion SLD. The two figures
below quantify this gap on the observed data alone (no model), and so
motivate why the joint architecture needs the multistate hazard channel
(Section~\ref{sec-multistate}) on top of the tumour-burden trajectory.

\begin{figure}

\centering{

\pandocbounded{\includegraphics[keepaspectratio]{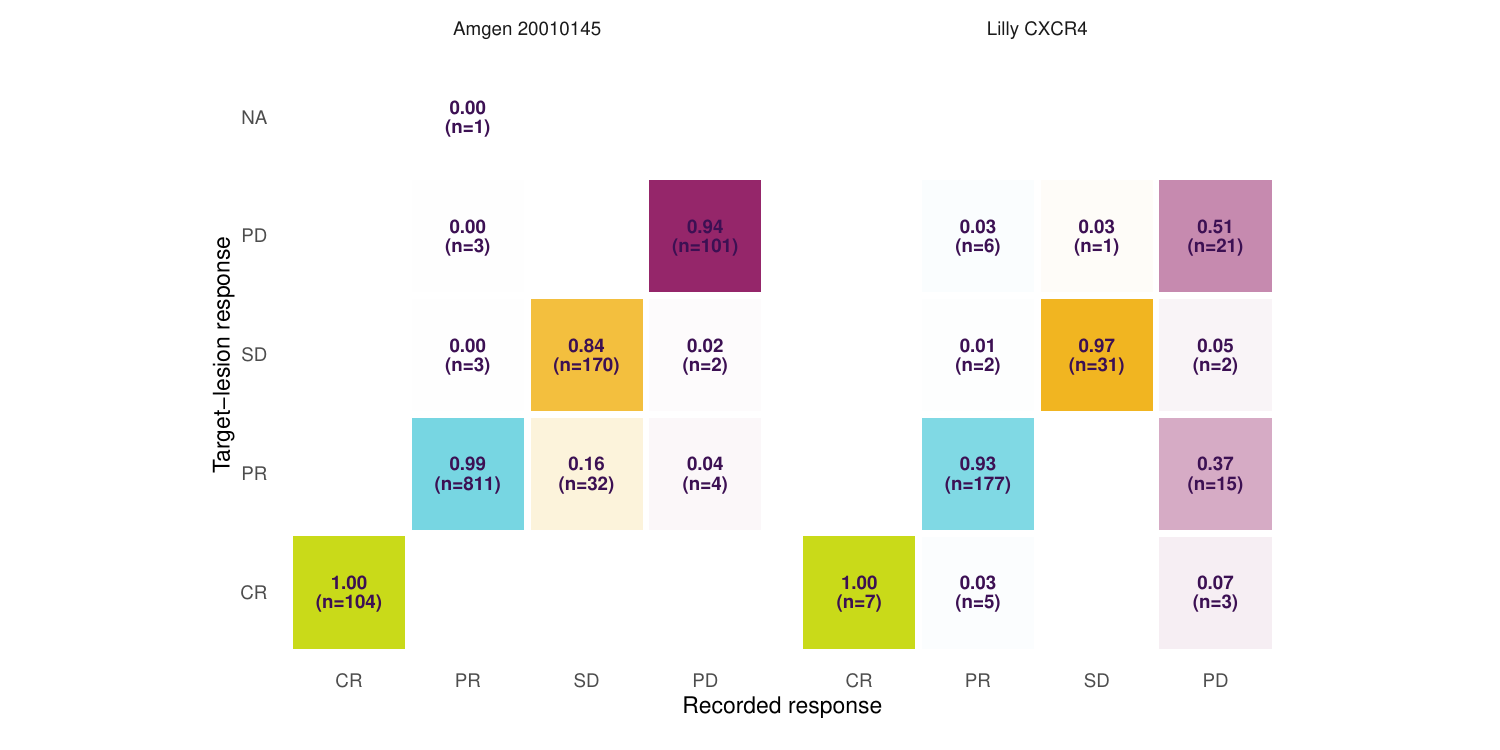}}

}

\caption{\label{fig-recist-confusion-sclc}\textbf{SCLC: recorded vs
target-lesion RECIST response.} Per-visit agreement between the recorded
clinical RECIST classification and the category computed from
target-lesion SLD alone. Cell shading is the row-normalised proportion;
off-diagonal cells are discordances driven by non-target/new lesions.}

\end{figure}

Figure~\ref{fig-recist-confusion-sclc} compares the recorded clinical
RECIST classification at each visit against a deterministic RECIST
category computed from the target-lesion SLD alone. Off-diagonal mass is
the response signal that target-lesion dynamics cannot reproduce: most
commonly, a recorded PD while the target SLD has not crossed the +20\%
progression threshold.

\begin{figure}

\centering{

\pandocbounded{\includegraphics[keepaspectratio]{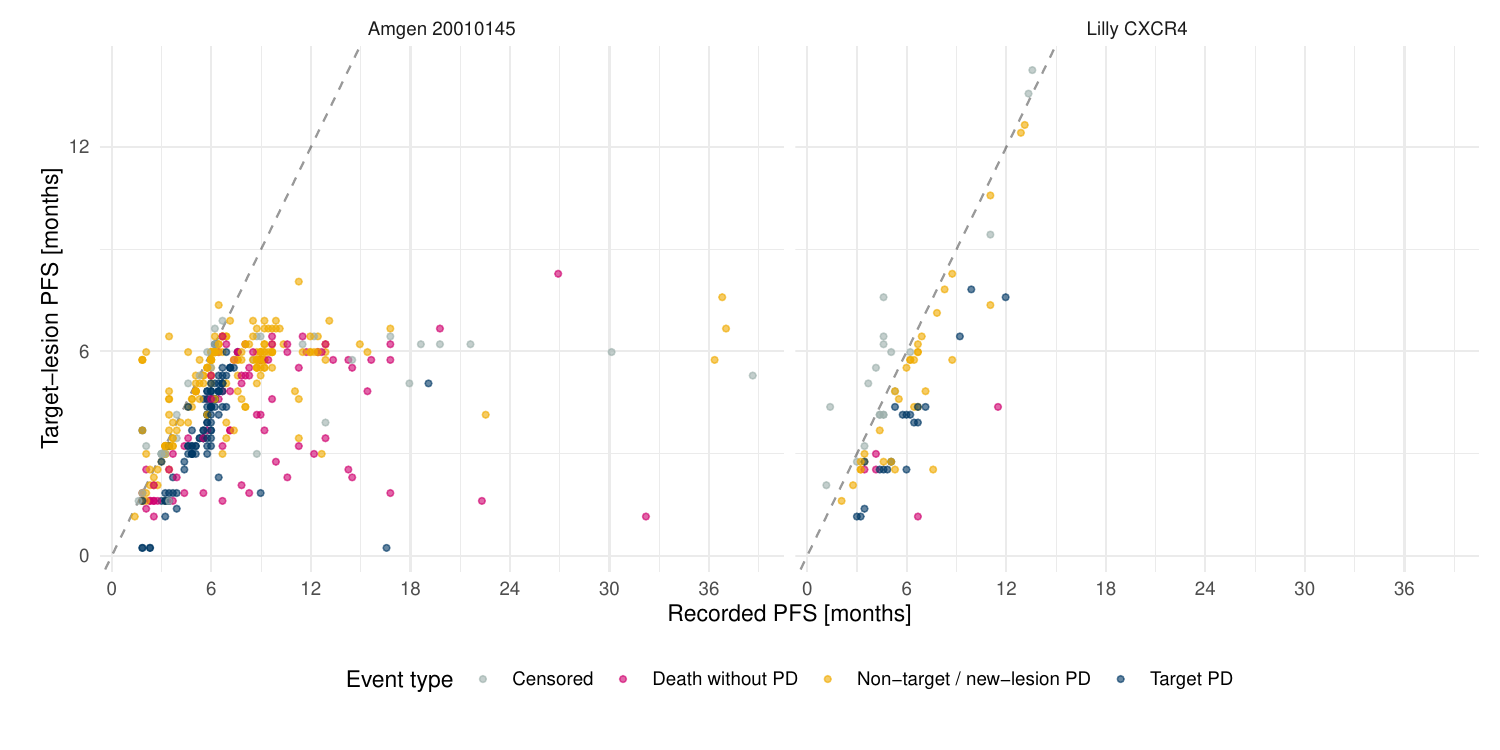}}

}

\caption{\label{fig-target-vs-recorded-pfs-sclc}\textbf{SCLC:
target-lesion PFS vs recorded clinical PFS.} Each point is a patient;
the dashed line is equality. Points below the diagonal: the clinical PFS
event was death or non-target/new-lesion PD while the target-lesion SLD
never crossed the progression threshold. Colour denotes event type; see
legend.}

\end{figure}

Figure~\ref{fig-target-vs-recorded-pfs-sclc} makes the same point at the
patient level for PFS: each point compares a patient's recorded clinical
PFS against a deterministic target-lesion PFS (the week the target SLD
first crosses the +20\% threshold). Points on the diagonal are
progression events the target lesion captures; points well below it are
clinical events --- death without progression, or non-target/new-lesion
PD --- that occur while the target SLD never crosses threshold. The
vertical spread below the diagonal is the share of the PFS estimand
carried by the hazard channel rather than the tumour-burden trajectory.

\protect\phantomsection\label{refs}

\end{document}